\documentclass[twocolumn]{aastex7}
\usepackage{amsmath, comment}
\shorttitle{Radiating Bondi Flows}
\shortauthors{Bailey et al.}
\received{\today}
\revised{\today}
\accepted{\today}
\submitjournal{MNRAS}

\begin{document}

\title{Radiating Bondi Flows I: Dimensionless Framework and Constant Opacity Solutions}

\author[orcid=0000-0002-6940-3161, gname=Avery, sname=Bailey]{Avery P. Bailey}
\affiliation{Steward Observatory, University of Arizona, 933 North Cherry Avenue, Tucson, AZ 85721-0065, USA}
\email[show]{averybailey@arizona.edu}

\author[0000-0002-3644-8726, gname=Andrew, sname=Youdin]{Andrew N.\ Youdin}
\affiliation{Steward Observatory, University of Arizona, 933 North Cherry Avenue, Tucson, AZ 85721-0065, USA}
\email{youdin@arizona.edu}

\author[0000-0001-5253-1338, gname=Kaitlin, sname=Kratter]{Kaitlin M. Kratter}
\affiliation{Steward Observatory, University of Arizona, 933 North Cherry Avenue, Tucson, AZ 85721-0065, USA}
\email{kkratter@arizona.edu}

\begin{abstract}
In this paper, we extend the physics of Bondi accretion to include the effects of radiative feedback in gas-pressure-dominated environments. We construct steady-state spherically symmetric accretion solutions including radiative heating and cooling. Under the simplifying assumption of a constant opacity, the solutions are controlled by four dimensionless parameters: the adiabatic index $\gamma$, optical depth through the Bondi radius $\tau_B$, dimensionless luminosity at infinity $\tilde{L}_\infty$, and a characteristic dimensionless cooling time $\beta$. We present numerical solutions across the dimensionless parameter space $(\tau_B, \tilde{L}_\infty, \beta)\in [10^{-3}, 10^3]$. Contrary to radiation-pressure-dominated environments, radiative feedback primarily operates to suppress accretion -- particularly at high $\tau_B$, $\tilde{L}_\infty$, and/or $\beta$. We also present analytic descriptions confirming the suppressive nature of this feedback and give the scalings for the accretion rate $\dot{M}\sim \tilde{L}_\infty^{-5/4}$ at large $\tilde{L}_\infty$, $\dot{M}\sim \tau_B^{-10/11}\beta^{-5/11}$ at large $\tau_B$, and $\dot{M}\sim (\tilde{L}_\infty\tau_B)^{-5/8}$ for large $\tilde{L}_\infty\tau_B$. We discuss the potential role of convection in these steady-state solutions, and the particular relevance to problems of planet formation where radiative heating is significant, but the system remains in the gas-pressure-dominated regime.
\end{abstract}



\section{Introduction}
Under the assumption of spherical symmetry and adiabatic energetics, \citet{Bondi1952} derived a fundamental steady-state accretion solution with wide-ranging astrophysical applications. Despite its simplicity, the original results of \citet{Bondi1952} remain remarkably powerful, relevant, and informative even some seven decades later. Nevertheless, in the interceding years, numerous works have improved upon the simple thermodynamics of \citet{Bondi1952} by including the physics of radiative energy transport. Primarily motivated by black hole accretion, works like \citet{KafkaMeszaros1976, Begelman1978, Thorne+1981, Flammang1982, Flammang1984} addressed optically thick cases through a diffusion approximation. These works determined that the accretion is indeed adiabatic for relativistic or sufficiently optically thick regimes. For non-relativistic regimes with lower optical depths however, the accretion becomes super-adiabatic, with accretion rate scaling inversely to optical depth $\dot{M}\sim \tau_B^{-1}$. At the same time, works like \citet{Shapiro1973} and \citet{Soffel1982}, concerned with optically thin or intermediary regime black hole accretion present rates consistent with Bondi's original calculation. 

Analogous extensions exist for various other astrophysical systems of interest -- optically thin galaxy cluster environments \citep{MathewsGuo2012}, neutron stars \citep{Maraschi+1978}, stars in AGN disks \citep{Chen+2024}, et cetera -- all adding to a patchwork of radiating Bondi\footnote{In this work, ``Bondi" as an adjective is used as shorthand for ``spherically symmetric steady-state''} solutions. But these environments, or at least the adopted fiducial conditions, are overwhelmingly those in which thermal pressure is a sub-dominant component of the energy density. 

There are works concerned with steady-state spherically symmetric accretion in thermally dominant regimes like star \& planet forming environments. See, for example, \citet{Stahler+1980,Stahler+1981,Stahler+1986,Omukai+2003,Hosokawa+2009}. But the associated strategy is to prescribe an accretion rate, decoupling the exterior flow dynamics from the radiative structure. This is useful for capturing the effects of self-gravity ($\dot{M}$ may be set to gravitational collapse solutions) and studying the interior structure of the flow in great detail, but it is distinct from solution of a global Bondi problem for $\dot{M}$.

The thrust of this work, is to advance the theory of non-adiabatic Bondi accretion for the less-developed gas-pressure-dominated regime, as has been done for the radiation-pressure dominated regime. In particular, this work is primarily interested in accretion rates and modification to the adiabatic accretion rate with the inclusion of radiative energy transport. As such, the main scientific deliverable is computation of a correction factor $f_{\rm acc}\equiv \dot{M}/\dot{M}_{\rm ad}$, measuring the radiative Bondi accretion rate relative to the equivalent adiabatic Bondi rate, as a function of environmental conditions.

This paper forms the first in a series on radiating Bondi solutions. Since the gas-pressure-dominated regimes have been relatively neglected, we find it prudent to make this paper as general as feasibly possible and concerned with a heavily theoretical dissection of this problem as an idealized ($r\rightarrow \infty$, constant opacity, etc.) mathematical one. This paper then, is in the spirit of established theoretical works like \citet{Bondi1952, Begelman1978, Thorne+1981, Flammang1982, Flammang1984}. Future papers will focus on particular contemporary applications -- treating them with the necessary specificity, and testing the idealizations made here by replacing them with prescriptions more suitable to the application at hand. For example, a subsequent Paper II examines in detail the application to the runaway phase of giant planet accretion with realistic opacities, dedicated hydrodynamic simulations, and interaction with relevant planetary processes like gap-opening. 

The structure of this paper is as follows. In Section \ref{sec:methods}, we briefly review the problem of adiabatic Bondi accretion before extending the formalism to include radiative transfer and details of our method of solution. Section \ref{sec:models} presents constant opacity radiating Bondi solutions and derives analytic scaling laws for different regions of parameter space. In Section \ref{sec:app}, we describe the applicability of these models. This includes the stability of solutions to convection, appropriate boundary conditions for physical problems. Finally, we summarize our findings and remark on potential applications in Section \ref{sec:end}.

\section{Problem Formulation and Method of Solution}\label{sec:methods}
\subsection{Classical Bondi Accretion}
\begin{figure}
    \centering
    \includegraphics[width=\linewidth]{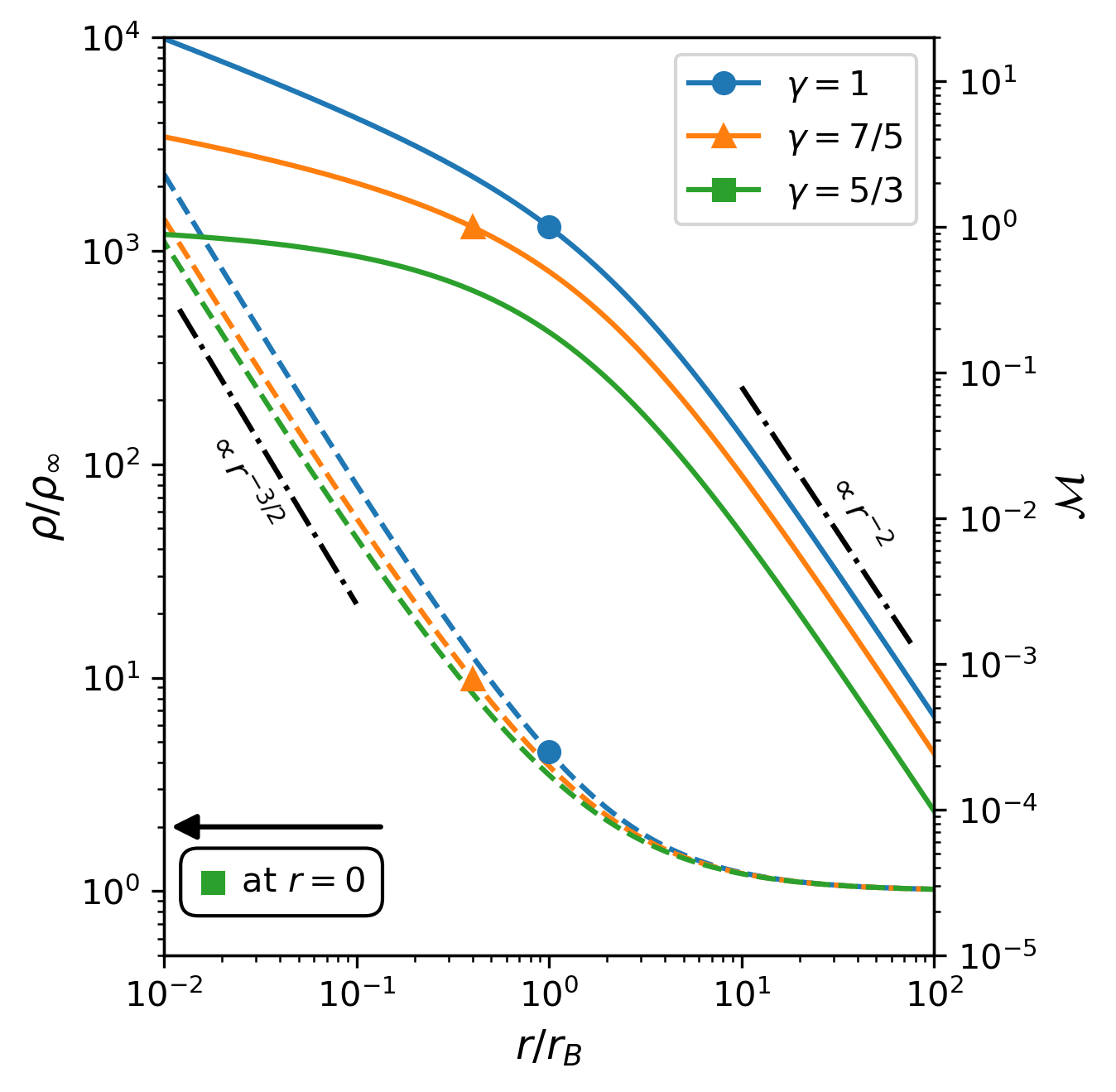}
    \caption{Adiabatic Bondi solutions for $\gamma=1$ (circle), $\gamma=7/5$ (triangle), $\gamma=5/3$ (square). The solid lines show Mach number profiles with corresponding density profiles as dashed lines. Markers are placed at the location of the sonic point with the $\gamma=5/3$ sonic point located at $r=0$. Dot-dashed lines plot the standard expected asymptotic scalings for reference.}
    \label{fig:adbondi}
\end{figure}

Given some ambient medium characterized by density, temperature at infinity $(\rho_\infty, T_\infty)$, Bondi accretion \citep{Bondi1952} is the spherically symmetric steady-state solution onto a gravitating body with mass $M$ obtained by solving the mass and momentum equations
\begin{subequations}
\begin{equation}\label{eq:mass}
\vec{\nabla}\cdot (\rho \vec{v}) = 0
\end{equation}
\begin{equation}
\vec{v}\cdot \nabla \vec{v} + \frac{\vec{\nabla} p}{\rho} = -\frac{GM}{r^2}\hat{r}.
\end{equation}
\end{subequations}
Classically these are closed with an equation of state for pressure $p \propto \rho^\gamma$ with $\gamma$, the ratio of specific heats, being bounded from below by $1$ (for an isothermal medium) and from above by $\gamma=5/3$. Under spherical symmetry $\vec{v}\rightarrow -v\hat{r}$ (and defining sound speed $c_s$, mach number $\mathcal{M}\equiv v/c_s$), these equations can be made more amenable to numerical solution as a set of coupled first-order ordinary differential equations e.g.\ \citet{ChoksiChiang2024}:
\begin{subequations}\label{eq:adiabatic_eqns}
\begin{equation}\label{eq:clrho}
\partial_r \ln\rho = \frac{2}{\gamma+1} \left(-\frac{2}{r} - \partial_r \ln \mathcal{M}\right)
\end{equation}
\begin{equation}\label{eq:clmach}
\partial_r \ln \mathcal{M}  = \frac{(\gamma+1)( 2c_s^2r - GM)}{2c_s^2 r^2\left(\mathcal{M}^2-1\right)} + \frac{\gamma-1}{r}
\end{equation}
\end{subequations}
While there exists a singularity at $\mathcal{M}=1$ when the denominator in (\ref{eq:clmach}) becomes zero, there exists regular solution (the Bondi accretion solution) when the numerator also vanishes and the solution passes through critical point at $r=GM/2c_s^2$, $\mathcal{M}=1$. These equations can be rescaled to depend only on a single free parameter -- the adiabatic index $\gamma$. For reference, the adiabatic Bondi solutions subject to common choices of adiabatic index $\gamma$ are provided in Figure \ref{fig:adbondi}.
The resulting transonic solutions have accretion rate 
\begin{equation}\label{eq:adrate}
    \dot{M}_{\rm ad} = \frac{\pi q_\gamma G^2M^2\rho_\infty}{c_{s,\infty}^3},
\end{equation}
depending weakly on $\gamma$ through the sound speed at infinity $c_{s,\infty}$ and prefactor
\begin{equation}
    q_\gamma(\gamma) = \left(\frac{2}{5-3\gamma}\right)^{(5-3\gamma)/(2\gamma-2)}
\end{equation}
with minimum $q_\gamma(5/3) = 1$, $q_\gamma(7/5) = 5/2$ and maximum $q_\gamma = e^{3/2}\approx 4.48$ for $\gamma=1$. The location of the corresponding critical point, the sonic radius, is given by
\begin{equation}
    r_{\mathcal{M}} = \frac{5-3\gamma}{4}\frac{GM}{c_{s,\infty}^2}
\end{equation}
which asymptotes to zero as $\gamma\rightarrow 5/3$. Various works will often utilize a Bondi radius $r_B\sim GM/c_{s,\infty}^2$ which may or may not be coincident with the true sonic radius of a flow. Similarly, the prefactor in the accretion rate is often dropped in favor of something like $\dot{M}\sim \rho_\infty r_B^2c_{s,\infty}$ \citep{Mordasini+2012, Emsenhuber+2021, Choksi+2023}. In this work, we define $r_B \equiv GM/2c_{s,\infty}^2$. This choice of $r_B$ is only equivalent to the true sonic radius in the isothermal $\gamma=1$ case. Because our focus is on the radiative aspects of the problem, in this work, we will generally fix $\gamma$ to an intermediate value ($\gamma=7/5$) where the distinction between $r_B$, $r_{\mathcal{M}}$ is not so great. With these definitions, $\gamma = 7/5$ gives:
\begin{align}
\dot{M}_{\rm ad} &= 10 \pi \rho_\infty c_{s,\infty} r_B^2 \, .
\end{align}
We also present accretion rates relative to adiabatic such that they can be applied without much concern for the choice of normalization of the Bondi rate.  

\subsection{Bondi Accretion with Radiative Heating \& Cooling}\label{sec:rad}
In the classical Bondi problem, the barotropic equation of state enforces a constant entropy profile ($\partial_r S = 0$) for the gas. In general however, variable entropy will alter the density/sound speed, modifying equations (\ref{eq:clrho}) \& (\ref{eq:clmach}). Utilizing the specific heat at constant volume $c_V$ to define the dimensionless entropy $s\equiv S/c_V$, equations (\ref{eq:clrho}) \& (\ref{eq:clmach}) become:
\begin{equation}\label{eq:rho}
\partial_r \ln\rho = \frac{2}{\gamma+1} \left(-\frac{2}{r} - \partial_r \ln \mathcal{M} - \frac{\partial_r s}{2}\right)
\end{equation}
\begin{equation}\label{eq:mom}
\begin{split}
\partial_r \ln \mathcal{M}  = & \frac{(\gamma+1)( 2c_s^2r - GM - c_s^2r^2\partial_r s/\gamma)}{2c_s^2 r^2\left(\mathcal{M}^2-1\right)}\\
&+ \frac{\gamma-1}{r} -  \frac{\partial_r s}{2}
\end{split}
\end{equation}

The entropy change along the flow, $\partial_r s$, arises from the balance of heating and cooling, which we take to be radiative.\footnote{See Section \ref{sec:convect} for our analysis of the role of convective heat transport.}  In our steady state flow, entropy changes as $Ds/Dt = -v \partial_r s$, making the energy equation
\begin{align}\label{eq:AYs}
\partial_r s &= \frac{1}{c_V T \dot{M}} \partial_r L   
\end{align} 
in terms of the luminosity, $L$.  We express this equation in terms of our preferred variables in Eq.\ (\ref{eq:s}).

Various radiative transfer schemes can be used to compute the net cooling $\partial_r L = 4 \pi r^2 \nabla \cdot \vec{F_r}$, from the radiative flux $\vec{F_r}$ ($= F_r \hat{r}$ in 1D). A general frequency-dependent scheme would express the flux as
\begin{equation}\label{eq:nuent}
\vec{\nabla}\cdot \vec{F}_r =\frac{1}{r^2}\partial_r (r^2 F_r)= \int4\pi\rho\kappa_\nu (S_\nu-J_\nu) d\nu,
\end{equation}
in terms of opacity $\kappa$, source function $S$, and radiation mean intensity $J$, with $\nu$-subscripts denoting frequency-dependent quantities. In principle, this 1D system can be closed with a form for $S_\nu$ and $J_\nu$. To address the former, we assume the source function to be described by the Planck function. For the latter, in principle, $J_\nu$ could be obtained from a solution of the transfer equation $\hat{n}\cdot \nabla I_\nu = 4\pi \rho \kappa_\nu(S_\nu-I_\nu)$ for intensity $I_\nu$ along direction $\hat{n}$. In this work, however, we make some simplifications to streamline solution of the equations. First we replace equation (\ref{eq:nuent}) with an approximate frequency-integrated form:
\begin{equation}\label{eq:flux}
\frac{1}{r^2}\partial_r (r^2 F_r) = 4\pi\rho\kappa_P(B - J)
\end{equation}
Now $B\equiv a_r c T^4/4\pi$ is the frequency-integrated Planck function with speed of light $c$, radiation constant $a_r$. $J$ is an approximate frequency-averaged mean intensity and $\kappa_P$ is the corresponding frequency-averaged (Planck) opacity. In principle, $J$ could be paired with a frequency-averaged opacity separate from $\kappa_P$, but since any frequency-averaged opacity for $J$ will be approximate if not corrected via iterative techniques, we use the standard $\kappa_P$ choice \citep{MihalasMihalas1984}. 

Expressing $J$ as a radiation energy density $E_r=4\pi J/c$, we then supplement the flux equation with a higher order moment equation for the radial component of the radiation pressure tensor $P_r$ under spherical symmetry:
\begin{equation}\label{eq:pr}
\partial_r P_r + \frac{3P_r - E_r}{r}= -\frac{\rho \kappa_R F_r}{c}
\end{equation}
and close the radiation subsystem with an appropriate closure relation \citep{Shu1992}:
\begin{equation}\label{eq:closure}
E_r = 3P_r - 2F_r/c \ .
\end{equation}
In this case, the opacity $\kappa_R$ is taken to be the Rosseland mean opacity so that the appropriate flux is recovered in the diffusion regime \citep{MihalasMihalas1984}. 

For integrating the system of differential equations (\ref{eq:rho}), (\ref{eq:mom}), (\ref{eq:AYs}), (\ref{eq:flux}), (\ref{eq:pr}) we prefer to use the dependent variables $(\rho, \mathcal{M}, s, L, E_r)$. In terms of our preferred variables, the full system of steady-state radiating Bondi equations to be solved in the following sections is then specified by equations (\ref{eq:rho}) \& (\ref{eq:mom}) along with the following three equations for the radiative transfer:
\begin{align}
&\partial_r s = \frac{\gamma(\gamma-1)}{\rho c_s^3\mathcal{M}}\frac{\partial_r L}{4\pi r^2} \label{eq:s} \\
& \partial_r L = 4\pi r^2\rho\kappa_P c(a_r T^4 - E_r)\\
&\partial_r E_r = -\frac{L}{4\pi r^2 c}\left(\frac{2}{r} + \frac{2\partial_r L}{L} + 3\rho \kappa_R\right)\label{eq:er}.
\end{align}
Closing the system requires expressing $T$, $c_s$ in terms of the dependent variables i.e.\ an equation of state. We adopt an ideal gas equation of state $p=\rho k_bT/\mu m_p$ with $k_B$ the Boltzmann constant and $\mu$ the mean molecular weight in units of proton mass $m_p$. This translates to
\begin{equation}
T = \frac{\mu m_pc_s^2}{\gamma k_B},
\end{equation}
\begin{equation}
c_s^2 = c_{s,\infty}^2 e^{s-s_\infty} \left(\frac{\rho}{\rho_\infty}\right)^{\gamma-1},
\end{equation}
using the thermodynamic state of the gas at infinity as a reference state.  The scaled entropy $s_\infty$ is an arbitrary reference point, not an independent free parameter.

In keeping with the classical Bondi formalism, we introduce the luminosity and radiation energy density at infinity $L_\infty$, $E_{r,\infty}$. The analogous radiating Bondi solution then, is the unique transonic solution to these equations, subject to the condition $(\rho, \mathcal{M}, s, L, E_r)\rightarrow (\rho_\infty, 0, s_\infty, L_\infty, E_{r,\infty})$ as $r\rightarrow \infty$.

Integrating these radiative Bondi equations requires specifying the parameters $\gamma$, $M$, $\mu$, $\rho_\infty$, $T_\infty$, $L_\infty$, $E_{r,\infty}$ and the opacity law functions $\kappa_R(\rho, T)$, $\kappa_P(\rho, T)$. Note that $E_{r,\infty}$ is not an independent free parameter, since radiative equilibrium, with $E_{r,\infty} = a_rT_\infty^4$ holds at infinity due to vanishing flow speeds.

\subsubsection{Dimensionless Framework}
 To reduce the number of free parameters, and develop a general understanding of radiative Bondi accretion, this work considers a simplified model with a single constant opacity $\kappa = \kappa_R = \kappa_P$.  

In dimensionless form, the standard adiabatic Bondi problem depends on a \textit{single} free  parameter $\gamma$.  We now show that our radiative Bondi problem depends on an additional \textit{three} dimensionless parameters; $\tau_B, \beta,$ and $\tilde{L}_\infty$, representing optical depth, scaled cooling time and scaled luminosity values.

We transform to dimensionless variables:
\begin{align}
        \tilde{r} &\equiv \frac{r}{r_B} \  & \tilde{\rho}&\equiv\frac{\rho}{\rho_\infty}  & \tilde{c}_s&\equiv \frac{c_s}{c_{s,\infty}}\\
        \tilde{T}&\equiv\frac{T}{T_\infty} & \tilde{E}_r &\equiv \frac{E_r}{a_rT_\infty^4} & \tilde{L}&\    \equiv \frac{L}{L_B} \nonumber
\end{align}
where $r_B\equiv GM/2c_{s,\infty}^2$ is the Bondi radius and $L_B\equiv 4\pi r_B^2a_rcT_\infty^4$ is a characteristic luminosity scale. With these variables, the dimensionless equation of state is
\begin{equation}
\tilde{T}=\tilde{c}_s^2 =e^{s-s_\infty}\tilde{\rho}^{\gamma-1}
\end{equation}
and the constant scaled accretion rate is
\begin{align}\label{eq:facc}
f_{\rm acc} &= \frac{\dot{M}}{\dot{M}_{\rm ad}}= \frac{\tilde{\rho}\sqrt{\tilde{T}}\mathcal{M}\tilde{r}^2}{q_\gamma}. 
\end{align} 

The non-dimensionalized governing equations are
\begin{subequations}\label{eq:dimless}
\begin{align}
&\partial_{\tilde{r}} \ln\tilde{\rho} = \frac{2}{\gamma+1}\left(-\frac{2}{\tilde{r}} - \partial_{\tilde{r}} \ln \mathcal{M} - \frac{\partial_{\tilde{r}} s}{2}\right) \label{eq:dim-rho}\\
&\begin{aligned}
\partial_{\tilde{r}} \ln \mathcal{M}  = &  \frac{(\gamma+1)(\tilde{c}_s^2\tilde{r} - 1 - \tilde{c}_s^2\tilde{r}^2\partial_{\tilde{r}} s/2\gamma)}{\tilde{c}_s^2 \tilde{r}^2\left(\mathcal{M}^2-1\right)} \label{eq:dim-M} \\
&+ \frac{\gamma-1}{\tilde{r}} -  \frac{\partial_{\tilde{r}} s}{2}
\end{aligned}\\
&\partial_{\tilde{r}} s = \frac{1}{4q_\gamma f_{\rm acc} \beta \tau_B}
\frac{\partial_{\tilde{r}} \tilde{L}}{\tilde{T}}\label{eq:dim-s}\\
& \partial_{\tilde{r}} \tilde{L} = \tilde{r}^2\tilde{\rho}
\tau_B (\tilde{T}^4 - \tilde{E}_r) \label{eq:dim-L}\\
&\partial_{\tilde{r}} \tilde{E}_r = -\frac{\tilde{L}}{\tilde{r}^2}\left(\frac{2}{\tilde{r}} + 2\partial_{\tilde{r}} \ln\tilde{L} + 3\tilde{\rho} \tau_B
\right)\label{eq:dim-er}
\end{align}
\end{subequations}
subject to the boundary condition $(\tilde{\rho}, \mathcal{M}, s, \tilde{L}, \tilde{E}_r)\rightarrow (1,0,s_\infty,\tilde{L}_\infty, 1)$ as $r\rightarrow \infty$.    

These equations, and the boundary condition on $\tilde{L}$ contain the four dimensionless free parameters that define a problem:
\begin{itemize}
    \item the ratio of specific heats $\gamma$
     \item a dimensionless luminosity at infinity $\tilde{L}_\infty \equiv L_\infty/L_B$.
    \item a characteristic optical depth  $\tau_B \equiv  \kappa\rho_\infty r_B$
    \item a dimensionless cooling time
    \begin{align}\label{eq:betacool}
\beta & \equiv  \frac{1}{4 \gamma (\gamma -1)} \frac{\rho_\infty c_{s,\infty}^3}{a c T_\infty^4} \frac{1}{\tau_B} = \frac{t_{\rm cool} c_{s,\infty}}{r_B} 
\end{align} 
   
\end{itemize}
The characteristic dimensional cooling time is  
\begin{align}
t_{\rm cool} &= \frac{c_V}{4 a c \kappa T_\infty^3} \, .
\end{align}
The role of $\beta$ as a cooling time is also seen by combining Eqs.\ (\ref{eq:dim-s}), (\ref{eq:dim-L}) as
\begin{align}
 \mathcal{M}\tilde{c}_s \cdot\tilde{T}\partial_{\tilde{r}} s  = 
 \frac{\tilde{T}^4 - \tilde{E}_r}{4 \beta}
\end{align}
to show that equilibrium solutions balance heat advection (at speed $\mathcal{M} \tilde{c}_s$) with radiative heating and cooling.

In the remainder of this work, we construct solutions to these dimensionless equations and generally present the results in terms of dimensionless parameters/variables so that they may be rescaled to any relevant astrophysical system. 

For context, we now consider the characteristic values of these dimensionless parameters, when applied to giant planet formation in a protoplanetary disk with a radial temperature and surface density profile, 
\begin{equation}
    T_\infty(R) = 100 \text{ K}\left(\frac{R}{10\text { AU}}\right)^{-1/2}
\end{equation}
\begin{equation}
    \Sigma_\infty(R) = 70 ~{\rm \frac{g}{cm^2}}\left(\frac{R}{10\text { AU}}\right)^{-3/2}
\end{equation}

In these environments (evaluating gas pressure $P_\infty$ at the disk midplane for a star of mass $M_\ast$), radiation pressure is indeed sub-dominant
\begin{equation}
    \frac{a_r T_\infty^4}{P_\infty} \approx 5\times 10^{-5}\left(\frac{R}{10\text { AU}}\right)^{5/4}\left(\frac{M_\ast}{M_\odot}\right)^{-1/2}
\end{equation}
consistent with our neglect of radiation pressure forces.  In more detail, we can express the condition $L_\infty \ll L_{\rm edd}$, the Eddington luminosity, as $\tilde{L}_\infty  \tau_B \ll 2 \rho_\infty c_{s,\infty}^2 / (a_r T_\infty^4)$ or, equivalently (for $\gamma = 7/5$) $\tilde{L}_\infty \ll 4.5 \beta c/c_{s,\infty}$.  We find that these criteria are satisfied in planet formation problems (see discussion below of expected $\tilde{L}_\infty$ values). However, these criteria for neglecting radiation pressure forces should be checked for other applications.

At low temperatures, mean opacities are dominated by dust, and are density-independent. For a dust-to-gas ratio of $1/100$ and a dust size distribution with power-law index $q=3.5$, and maximum particle size of 1 cm, the Rosseland mean opacity is approximately \citep{Birnstiel+2018, Zhu+2021}
\begin{equation}
    \kappa \approx 0.4 ~\frac{\text{ cm}^2}{{\rm g}}\left(\frac{T_\infty}{100 \text{ K}}\right)^{1/2}  .
\end{equation}
For a planet of mass $M$, the values of the dimensionless parameters then evaluate to
\begin{subequations}\label{eq:taubeta_vals}
\begin{equation}
\tau_B = 0.5\left(\frac{M}{10 M_\oplus}\right)\left(\frac{R}{10 \text{ AU}}\right)^{-5/2} \left(\frac{M_\ast}{M_\odot}\right)^{1/2}
\end{equation}
\begin{equation}
\beta  = 0.04\left(\frac{M}{10 M_\oplus}\right)^{-1}\left(\frac{R}{10 \text{ AU}}\right)\ .
\end{equation}
\end{subequations}
For different planet masses and formation locations, these optical depth values span thin and thick regimes, while $\beta$ tends to be less than unity.

The expected values of $\tilde{L}_\infty$ require further discussion. This work treats $\tilde{L}_\infty$ as a free parameter, so that we compute the accretion efficiency $\dot{M}/\dot{M}_{\rm ad} \equiv f_{\rm acc}(\tau_B,\tilde{L}_\infty, \beta)$ following the solution procedure in \S \ref{sec:ext}. Astrophysically, $\tilde{L}_\infty$ is not a free parameter -- it is necessarily related to the luminosity $L_{\rm src}$ at the surface of the accretor $r_{\rm src}$, i.e.\ the accretion luminosity $\sim GM\dot{M}/r_{\rm src}$ plus any intrinsic luminosity of the accretor (which we ignore in this discussion as sub-dominant). In principle, one would need to integrate solutions from $r\in [r_{\rm src}, \infty)$ to establish a relation between $L_{\rm src}\leftrightarrow \tilde{L}_\infty$ and apply our accretion rates for given $\tilde{L}_\infty$ to systems with physical radius $r_{\rm src}$. However, as we show in \S \ref{sec:freefall} and subsequent numerical integrations, an accretor with physical size smaller than the sonic radius  $r_{\rm src}\lesssim r_{\mathcal{M}}$ (a true Bondi solution), must necessarily have a luminosity profile that is nearly constant with radius $d\ln \tilde{L}/d\ln \tilde{r} < O(1)$. Therefore, for self-consistent radiative Bondi solutions considered here the approximation $L_{\rm src}\approx L_{\infty}$ made throughout this work is justified and no interior integration is necessary\footnote{We still perform interior integrations for the models presented in \S \ref{sec:models}, but this is for purely demonstrative purposes.}.

Even with the $L_{\rm src}\approx L_\infty$ approximation, a self-consistent value of $L_{\rm src}$ is not known a priori, since luminosity is a function mass accretion rate which itself depends on luminosity $L_{\rm src}\sim \dot{M}(L_{\rm src})$. In dimensionless terms,
\begin{equation}\label{eq:Linfacc}
  \tilde{L}_{\rm src}  = \frac{56}{5} \frac{r_B}{r_{\rm src}} \beta \tau_B f_{\rm acc}\,
\end{equation}
with $f_{\rm acc}(\tilde{L}_{\rm src})$ unknown. Determination of a self-consistent luminosity then requires solution of the implicit luminosity equation (\ref{eq:Linfacc}) either iteratively or a posteriori. We adopt the latter and compute a model grid of $f_{\rm acc}(\tau_B, \tilde{L}_\infty, \beta)$. Across this model grid, Eq.\ (\ref{eq:Linfacc}) (or some other well-motivated relation between $\tilde{L}_{\rm src}\leftrightarrow \dot{M}$) is applied after-the-fact, to select the model with self-consistent accretion luminosity given physical parameters like $r_{\rm src}$.

Paper II follows a similar strategy to determine self-consistent solutions (but relaxing the constant opacity assumption) for planet-forming scenarios.  For now we note that large values of $r_B/r_{\rm src}$ (planetary surface well inside the Bondi radius) often require suppressed accretion $f_{\rm acc} < 1$ to avoid too large values of $\tilde{L}_{\rm src}$.  This work will show that large $\tilde{L}_{\rm src}$ lowers $f_{\rm acc}(\tau_B,\tilde{L}_\infty, \beta)$ (see \S\ref{sec:num}). Paper II will show that  $\tilde{L}_{\rm src}\approx 1-10$ values are typical in consistent planetary accretion solutions.

\subsection{Numerical Solution of Steady State Equations}\label{sec:steady}

\subsubsection{Integration Exterior to the Sonic Point}\label{sec:ext}
The full structure of the accretion flow is not required to determine the accretion rate. Simply knowing the location of the sonic point ends up being sufficient, since $\dot{M}=4\pi r^2\rho v$ is a constant in steady-state and may be evaluated anywhere in the flow. 
Given boundary conditions at infinity, equations (\ref{eq:dim-rho})-(\ref{eq:dim-er}) can be integrated as an initial value problem. By iteratively adjusting the boundary conditions, the integrations can be made to converge to the critical curve, thus determining the location of the sonic point, the exterior flow structure, and the steady-state mass accretion rate.

Schematically our procedure to solve for accretion rates is as follows. We begin at a finite but suitably large radius $r_\infty$ exterior to the sonic point where four of the five dependent variables are sufficiently close to their values at infinity $(\rho_\infty, s_\infty, L_\infty, E_{r,\infty})$. We then guess a value for $\dot{M}$ at this exterior point (effectively giving the boundary condition on $\mathcal{M}$ through $\dot{M}=4\pi r^2\rho \mathcal{M}c_s$) and integrate inwards, categorizing the solution as one which either undershoots or overshoots the sonic point based on its trajectory. We repeat the integration, changing $\dot{M}$ and iterating on its value until the solution is brought sufficiently close to the sonic point to terminate the process. 
\begin{figure}
    \centering
    \includegraphics[width=\linewidth]{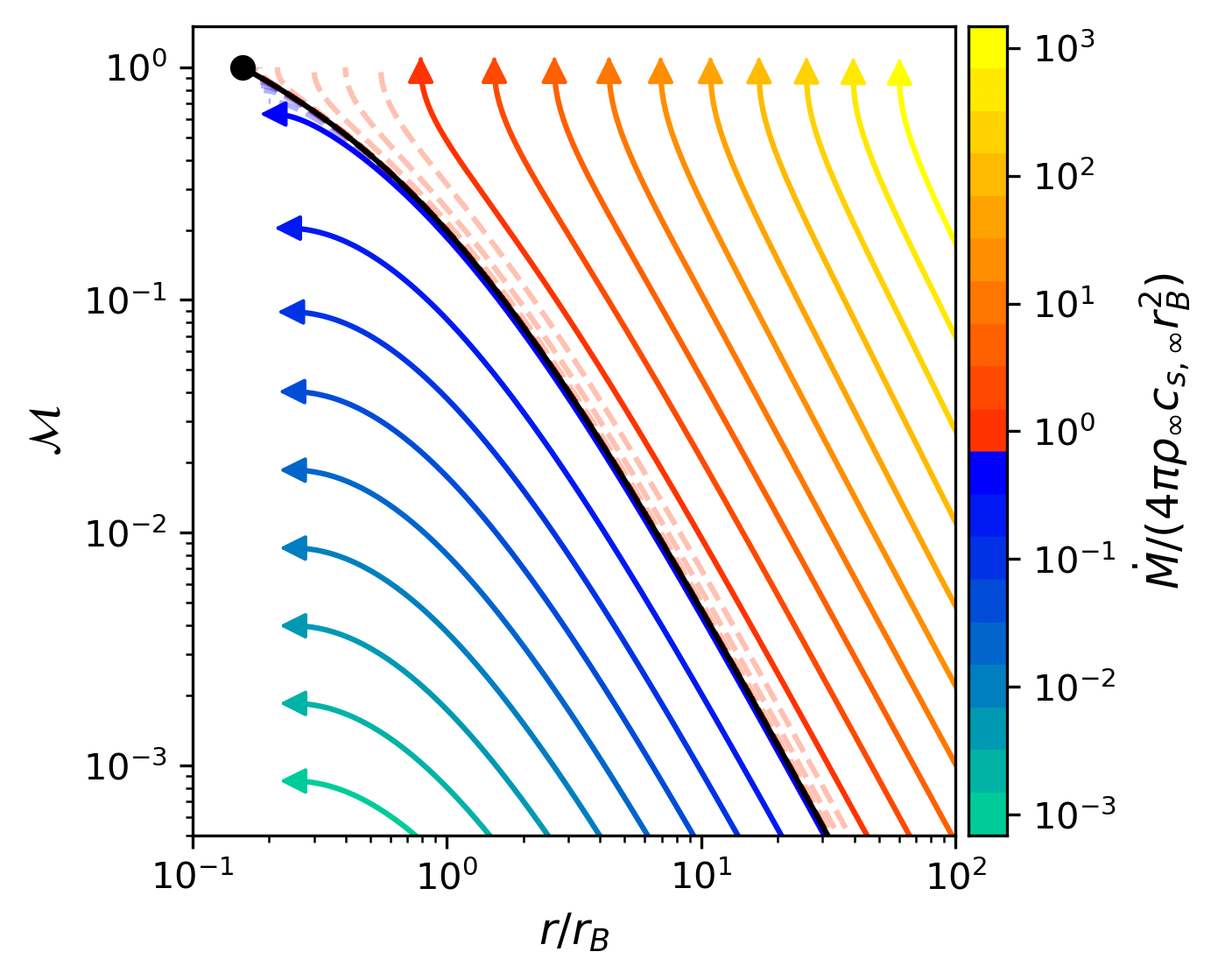}
    \caption{Example of families of curves and iterative procedure used to determine the sonic point (and therefore the accretion rate) for a model with $(\gamma, \tau_B,\tilde{L}_\infty, \beta) = (7/5,1,10,1)$. Each curve is an integration  colored according to the choice of mass accretion rate. Cool-hued curves belong to the family of solutions which undershoot the sonic point, while warm hues are solutions classified as overshooting. Faint dashed lines show the steps of our iterative search to determine the sonic point, with a solid black line marking the converged transonic solution. An adiabatic solution would have $\dot{M}=10\pi\rho_\infty c_{s,\infty}r_B^2$, or a value of $5/2$ in the presented colorbar. It is seen that such an accretion rate would overshoot the sonic point, and that radiative feedback skews the transonic solution inwards to a comparably lower accretion rate for this set of dimensionless parameters.}
    \label{fig:iterate}
\end{figure}
We plot an example of integrations under different $\dot{M}$ in Figure \ref{fig:iterate} and the results of our iterative process to determine the transonic solution. Curves which undershoot the sonic point belong to the family of solutions termed Type I in the original work of \citet{Bondi1952} while the family of curves that overshoot the sonic point are unphysical. Requiring the solution to be transonic and lie between these two families of curves, sets the condition on $\dot{M}$. 

While this procedure is robust, we find that outright integration of equations (\ref{eq:dimless}) is numerically problematic when the system is very close to radiative equilibrium -- as is the case at large radius. To get around this issue, we begin the integrations solving instead a system of radiative equilibrium equations (see Appendix \ref{app:inteq}) and switch to the more general disequilibrium system (\ref{eq:dimless}) once the equilibrium assumption is expected to fail or $r<10^2r_B$, whichever happens first. 

Our integrations utilize the $\texttt{LSODA}$ method of $\texttt{solve\_ivp}$ from the $\texttt{scipy}$ package and error tolerances $\texttt{rtol} = \texttt{atol} = 10^{-14}$. All integrations start from outer boundary at $r=10^{12}r_B$ and adopt a threshold to switch from the radiative equilibrium system of equations to the disequilibrium equations of $\delta E/E = 10^{-3}$ (see Appendix \ref{app:inteq}).

\subsubsection{Integration through the Sonic Point}
Outright integration of Eq.\ (\ref{eq:dim-M}) fails near the sonic point due to the singular $\mathcal{M}^2-1$ denominator. To overcome this difficulty we derive a non-singular quadratic expression for $\partial_{\tilde{r}}\ln\mathcal{M}$ (see Appendix \ref{app:quadratic}) which is exact at the sonic point $(r,\rho, \mathcal{M},s, L,E_r) = (r_\mathcal{M},\rho_\mathcal{M},1,s_\mathcal{M}, L_\mathcal{M}, E_{r,\mathcal{M}})$. After determining the critical steady-state exterior solution (\S \ref{sec:ext}, Figure \ref{fig:iterate}), we re-integrate, switching to this well-behaved form over a narrow window near the sonic point ($|\mathcal{M}-1|\lesssim 0.05\equiv \delta_\mathcal{M} $) using the local state variables in place of their exact sonic values e.g.\ $\rho(r)\approx\rho_\mathcal{M}$. This introduces an error of $O(\delta_\mathcal{M})$ in $\partial_{\tilde{r}}\mathcal{M}$ or $O(\delta_\mathcal{M}^2)$ in $\mathcal{M}$, similar to Taylor expansion around the sonic point \citep{Zytkow1972} of first-order. Integration of the interior solution is simply achieved by switching back to Eq.\ (\ref{eq:dim-M}) upon exiting the window.

\section{Radiative Accretion Rates}\label{sec:models}
\subsection{Numerical Solutions}\label{sec:num}

We now present the accretion solutions for our radiative model, using the methods  described above.  We present radiative accretion rates, $\dot{M}$ relative to the adiabatic case, as an accretion efficiency, $\dot{M}/\dot{M}_{\rm ad} \equiv f_{\rm acc}(\tau_B,\tilde{L}_\infty, \beta; \gamma)$ in terms of general dimensionless parameters.  However we fix $\gamma = 7/5$ to reduce dimensionality.  In the adiabatic case, variations of  $1\leq  \gamma < 5/3$ in the allowed range cause modest changes to $\dot{M}$.  Also, $\gamma = 7/5$ represents a relevant case for cool diatomic gas.

We quantify radiative effects with a logarithmic model-grid spanning $(\tau_B, \tilde{L}_\infty, \beta)\in[10^{-3}, 10^3]$ with 21 points per dimension.  By spanning large and small values of the governing radiative parameters, we aim for a complete description of the solution space.  Extrapolation to parameters that lie outside this finite range is aided by our analytic modeling in \S\ref{sec:anal}.

\begin{figure*}
    \centering
    \includegraphics[width=\linewidth]{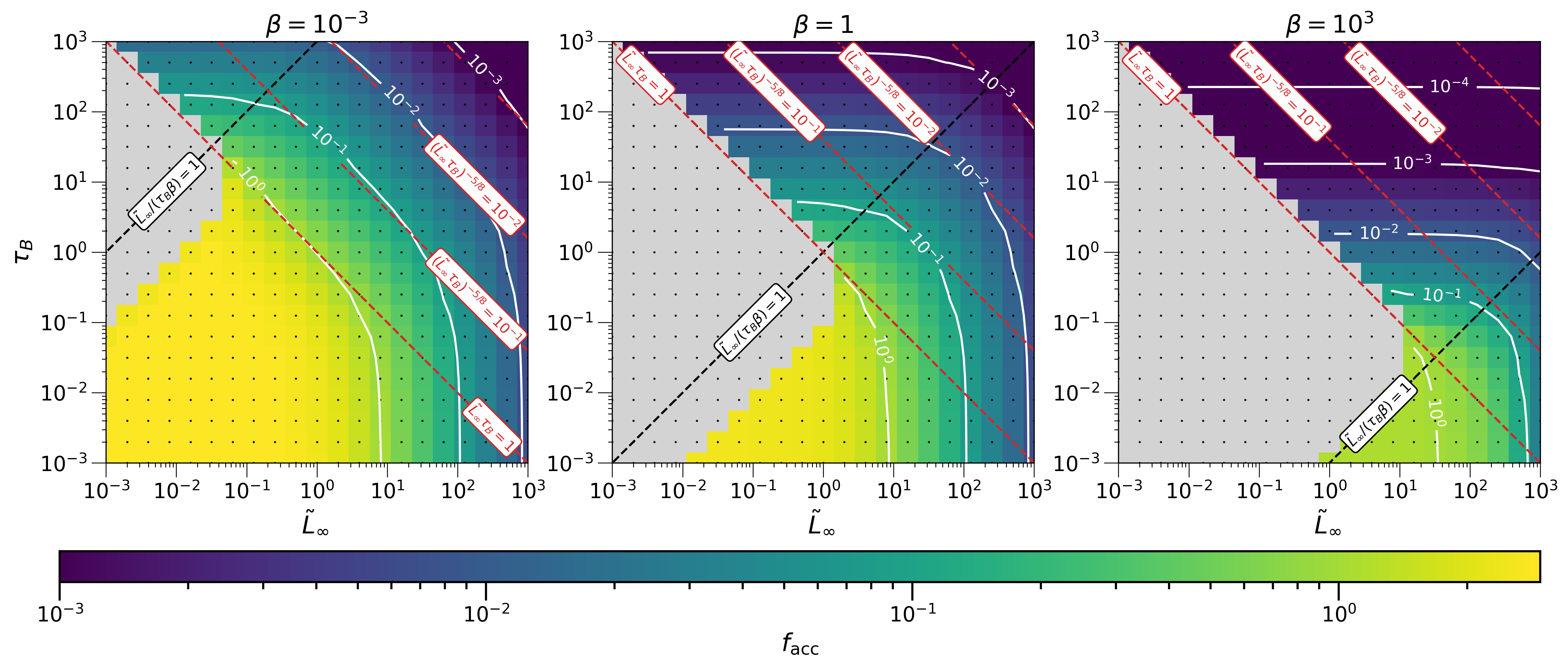}
    \caption{Slices at fixed $\beta\in(10^{-3},1, 10^3)$ in the surveyed parameter space, mapping the steady-state accretion rates relative to the adiabatic Bondi rate. Each black dot corresponds to a computed steady-state solution on our model grid with the surrounding square colored according to the value of $f_{\rm acc}$. Cells for which no transonic solution could be found are colored gray. Solid white contours are also drawn for each decade of $f_{\rm acc}$. For reference, dashed lines for the conditions $\tilde{L}_\infty = \tau_B\beta$ and $f_{\rm acc}=(\tilde{L}_\infty\tau_B)^{-5/8}$ with $f_{\rm acc} = (10^{-3}, 10^{-2}, 10^{-1}, 1)$ are included (see Section \ref{sec:anal} for the origin of these scalings).}
    \label{fig:facc}
\end{figure*}

The accretion rates in our model-grid computation are visualized in Figure \ref{fig:facc}, via slices at $\beta\in(10^{-3}, 1, 10^3)$. The full 3D grid of $f_{\rm acc}(\tau_B,\tilde{L}_\infty, \beta)$, including all $\beta$ values, is available for download.\footnote{https://github.com/apbailey/radiative-bondi-products}

Figure \ref{fig:facc} reveals several regimes of interest. First are blank regions of parameter space, occurring at $\tilde{L}_{\infty}/\beta \lesssim\tau_B\lesssim1/\tilde{L}_\infty$, where no transonic solution exists.\footnote{We do not consider subsonic, or ``Type I"  \citep{Bondi1952},  accretion solutions, as the required pressure support leads to implausibly large densities \citep{Shu1992}.}  At these low values of the escaping luminosity $L_\infty$, the solutions become un-physical, developing negative luminosities at smaller $r$. 

For low values of luminosity, optical depth and cooling time -- i.e.\ small $\tau_B$, $\tilde{L}_\infty$, $\beta$ but with $\tilde{L}_\infty \gtrsim \beta \tau_B$ for allowed solutions -- Figure \ref{fig:facc} shows that accretion rates approach the isothermal limit $f_{\rm acc}=\dot{M}_{\rm iso}/\dot{M}_{\rm ad} = 2(7e/5)^{3/2}/5\approx 2.97$.  Radiative accretion rates never exceed this isothermal, i.e.\ maximally cooling, limit.

Finally,  Figure \ref{fig:facc} shows substantial regions of parameter space -- at 
larger value of $\tau_B$, $\tilde{L}_\infty$, and/or $\beta$ -- where accretion is suppressed by radiative feedback.  The contours of constant $f_{\rm acc}$ reveal a complex parameter dependence to this suppression.  At large $\tau_B$, horizontal contours show where the suppressed accretion rate  is independent of luminosity. At large $\tilde{L}_\infty$, vertical contours show where $f_{\rm acc}$ becomes independent of the optical depth. For intermediate values of $\tilde{L}_\infty$, $\tau_B$, diagonal contours show a regime where $f_{\rm acc}\sim (\tilde{L}\tau_B)^{-5/8}$ scaling, modeled by the red contours in Figure \ref{fig:facc}. This intermediate scaling is most significant for rapid cooling (small $\beta$ values) where it gives a more extended transition between the large $\tau_B$ and large $\tilde{L}_\infty$ limits. To better understand these different regimes of radiative suppression, the following section develops analytic results for different limits of our full model.

\subsection{Analytic Description}\label{sec:anal}
\subsubsection{The Radiative Equilibrium Approximation}\label{sec:radeq}
While our full model includes different temperatures for matter and radiation, these temperatures are similar in much of parameter space, so that radiative equilibrium, with $\tilde{E}_r \simeq \tilde{T}^4$, holds to good approximation.  Using this approximation considerably simplifies the analysis. We first consider when radiative equilibrium is a good approximation.   Combining the energy equations (\ref{eq:dim-s}) \& (\ref{eq:dim-L}) gives the deviation from radiative equilibrium as
\begin{align}\label{eq:radeq_analysis}
\frac{\tilde{E}_r }{\tilde{T}^4} - 1  &= - 4 \beta q_\gamma\frac{d s}{d \ln \tilde{r}}   \frac{f_{\rm acc}}{\tilde{r}^3 \tilde{\rho} \tilde{T}^{3}} 
\end{align} 
using equation (\ref{eq:facc}).

The factors controlling radiative equilibrium are understood as follows.  The cooling time is  main dimensionless parameter controlling radiative equilibrium, with smaller $\beta$ values favoring equilibrium.  The entropy gradient determines whether accreting matter cools by radiating, with $ds/d\ln \tilde{r} > 0$ and  $\tilde{T}^4 > \tilde{E}_r$,  or is heated by radiation (vice versa).  Our solutions show that $|ds/d\ln \tilde{r}|$ values rarely exceed order unity.  Suppressed accretion, with smaller $f_{\rm acc}$ values, has lower flow speeds that favor radiative equilibrium.

At large radii, where $\tilde{\rho}\simeq \tilde{T} \simeq 1$, the $1/\tilde{r}^3$ factor in Eq.\ (\ref{eq:radeq_analysis}) strongly favors radiative equilibrium.  We exploit this fact to numerically extend our outer boundaries to large radii (Appendix \ref{app:inteq}).  The amount of radiative disequilibrium at smaller radii depends on the cooling time and the properties of the flow, $f_{\rm acc}$ and how $\tilde{r}^3 \tilde{\rho} \tilde{T}^{3} d \ln \tilde{r}/ds$ scales with radius.

Since these flow properties are difficult to predict in general, we define a local, positive definite measure of radiative disequilibrium,
\begin{equation}
    D_{\rm eq}\equiv \frac{|\tilde{E}_r - \tilde{T}^4|}{\min(\tilde{T}^4, \tilde{E}_r)}, 
\end{equation}
For each model in our 3D parameter space grid, we compute this $D_{\rm eq}$ measure as a function of radius along the flow, from the outer boundary to the sonic point.  The largest values occur near the sonic point. 

\begin{figure}
    \centering
    \includegraphics[width=\linewidth]{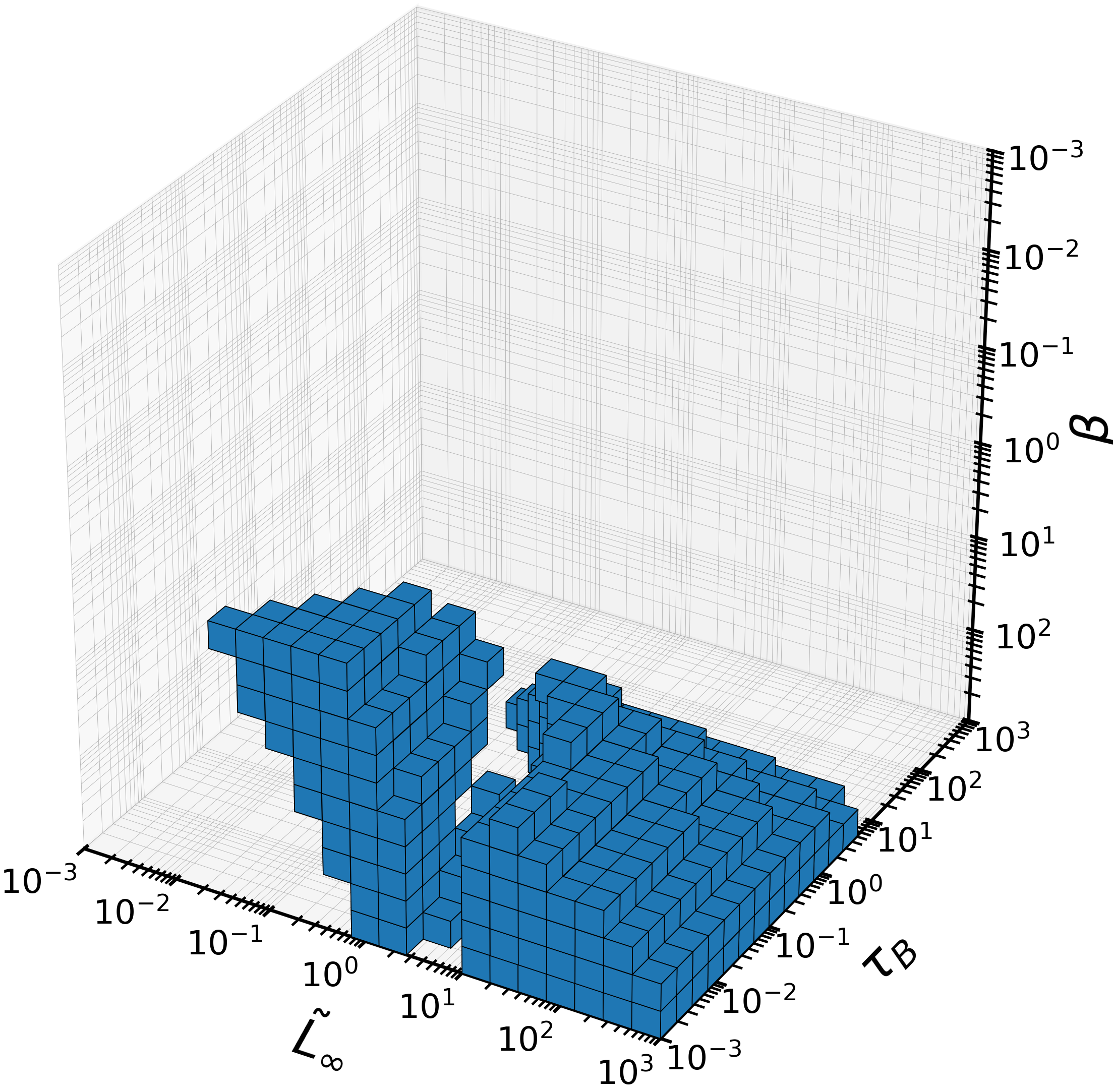}
    \caption{A volumetric rendering over the fiducial parameter space of our numerical solutions, showing the models which exhibit significant radiative disequilibrium $(E_r\neq aT^4)$. A blue cube is drawn at the location of any model which has $\max(D_{\rm eq}) > 1$. Note the z-axis is inverted for visualization purposes, so large values of $\beta$, favoring disequilibrium, are lower in a vertical sense.}
    \label{fig:diseq}
\end{figure}

Figure \ref{fig:diseq} shows which models have significant radiative disequilibrium, quantified as $D_{\rm eq} >1$ (order unity deviation) 
anywhere in the flow. Consistent with Eq.\ (\ref{eq:radeq_analysis}), radiative equilibrium is a good approximation for $\beta < 1$.  Larger optical depths generally favor radiative equilibrium.  Also  larger $\tau_B$ and $\tilde{L}_\infty$ values lower $f_{\rm acc}$ and favor radiative equilibrium, pushing the disequilibrium boundary to larger $\beta$. 

Altogether, the disequilibrium models plotted in Figure \ref{fig:diseq} amount to $\approx 5\%$ of our surveyed parameter space. If we relax the disequilibrium criterion to $\max(D_{\rm eq}) > 0.1$, this percentage increases to $\approx 15\%$. Thus arguments and scalings assuming radiative equilibrium will apply fairly well to the parameter space surveyed in Section \ref{sec:num} but caution is warranted for applying the results to systems at large $\beta$ and small $\tau_B$ where disequilibrium effects can arise.

We now consider a set of simplified model equations. First, note that the entropy equation (\ref{eq:dim-s}) can be rewritten in terms of a conserved total energy, or Bernoulli constant,
\begin{equation}\label{eq:etot}
    \partial_{r}\left[\dot{M}\left(\frac{v^2}{2}+\frac{c_s^2}{\gamma-1} - \frac{GM}{r}\right)-L\right]=0
\end{equation}
which does not vary across the flow. Making the approximation of radiative equilibrium $\tilde{E}_r=\tilde{T}^4$, then Eq.\ (\ref{eq:dim-er}) simply becomes a temperature equation. Because Eq.\ (\ref{eq:etot}) is outright integrable, the luminosity-temperature system in dimensionless form is:
\begin{subequations}\label{eq:reqTL}
\begin{equation}\label{eq:reqT}
   \partial_{\tilde{r}} \tilde{T}=-\frac{{\tilde{L}}}{4{\tilde{r}}^3\tilde{T}^3}\left(2C_1 + 3\tilde{\rho} \tau_B \tilde{r}
\right)
\end{equation}
\begin{equation}\label{eq:reqL}
    \tilde{L} = \tilde{L}_\infty + C_2 f_{\rm acc} \beta\tau_B\left[\frac{\mathcal{M}^2\tilde{T}}{2}
    +\frac{\tilde{T}-1}{\gamma-1} + \frac{2}{\tilde{r}^2}\right]
\end{equation}
\end{subequations}
where $C_1\equiv 1 +d\ln \tilde{L}/d\ln\tilde{r}$ is of order unity and  $C_2\equiv4\gamma(\gamma-1) q_\gamma $.

At large $\tilde{r}$ we may make a hydrostatic approximation ($\mathcal{M}^2\ll 1$) and drop the $\mathcal{M}^2$ term in Eq.\ (\ref{eq:reqL}). While this approximation breaks down at the sonic point, it holds reasonably well quite close to that point. At even larger radius, where gravity is weak ($\tilde{T}\gg 1/\tilde{r}^2 $), hydrostatic equilibrium may be further reduced to a condition of pressure equilibrium, $\partial_{\tilde{r}}\ln \tilde{\rho}=-\partial_{\tilde{r}}\ln\tilde{T}$. In this case, the $\sim 1/\tilde{r}^2$ term in Eq.\ (\ref{eq:reqL}) is also discarded and $\tilde{\rho}\sim1/\tilde{T}$. Under various levels of approximation, these coupled temperature, luminosity equations are sufficient to construct analytic scaling relations and understand the properties of models near radiative equilibrium as in the following sections.

\subsubsection{Free-fall solutions}\label{sec:freefall}
For  astrophysical accretion solutions, we need to validate the approximation that the escaping luminosity matches the accretion luminosity at the surface of the planet or other accretor, as in Eq.\ (\ref{eq:Linfacc}).  Here we justify this approximation inside the sonic point, with a general argument.  The following subsections consider specific accretion regimes, and validate constant $L$ outside $r_\mathcal{M}$ as well.  

We perform this analysis using scaled variables, and approximate order of magnitude arguments.  The sonic point is located where gravitational and thermal energies match, at a temperature $\tilde{T}_\mathcal{M} \sim 1/\tilde{r}_\mathcal{M}$.  
The condition for constant $L$ is approximately $d \ln \tilde{L}/d \ln \tilde{r} \lesssim 1$ or, from Eq.\ (\ref{eq:dim-s})  $\tilde{L} \gtrsim f_{\rm acc} \beta
\tau_B \tilde{T}$, since $ds/d\ln \tilde{r}$ is a (small) order unity quantity.  Note that this condition does not assume radiative equilibrium.  Applying this condition to the accretion luminosity, Eq.\ (\ref{eq:Linfacc}), and evaluating at the accretion surface (where $\tilde{T}$ is highest), gives $\tilde{T}(\tilde{r}_{\rm src}) \lesssim 1/\tilde{r}_{\rm src}$ as the condition for constant luminosity.

This simple condition requires that gravitational energy exceeds thermal energy near the surface.  This condition will be satisfied if the flow maintains supersonic free-fall.  In practice this condition requires that the accretion surface lies inside the sonic point $\tilde{r}_{\rm src} \lesssim \tilde{r}_\mathcal{M}$, as expected.  The constraint on $\tilde{T}(\tilde{r}_{\rm src})$ also requires a temperature profile that is shallower than $\tilde{T} \propto 1/\tilde{r}$. Both optically thick $(\sim r^{-5/8})$ and thin $(\sim r^{-1/2})$ transfer satisfy this condition.

In summary, constant $L$ inside the sonic point should be a good approximation for any standard accretion flow that remains supersonic.  The following subsections show how this condition gives specific constraints on the accretor size, $\tilde{r}_{\rm src}$.  

\subsubsection{No-Solution Regime}\label{sec:nosol}
We found in Section \ref{sec:num} that no accretion solutions exist between $\tilde{L}_\infty/\beta \lesssim \tau_B \lesssim 1/\tilde{L}_{\infty}$, because $d\tilde{L}/d\tilde{r} > 0$ leads (integrating inward) to negative luminosity before reaching the sonic point. 
We now explain these optically thin ($\tau_B \beta$) and thick ($1/\tau_B$) limiting luminosities, and why astrophysical accretion solutions avoid this no-solution space.

For the optically thin case, we make the isothermal approximation ($\tilde{T} \approx 1$) because the accretion solutions that neighbor this boundary are nearly isothermal, at least outside the sonic point.  We thus take $f_{\rm acc} \approx 1$ with the sonic point at $\tilde{r}_\mathcal{M}  \approx 1$.  With these approximations,  Eq.\ (\ref{eq:reqL}) gives $\tilde{L} > 0$ if $\tilde{L}_\infty \gtrsim \tau_B \beta /\tilde{r}$.  Thus positive luminosities down to the sonic point require $\tilde{L}_\infty \gtrsim \tau_B \beta$.    

This approximate argument reproduces the finding that solutions exist for $\tau_B \lesssim \tilde{L}_\infty /\beta$.  In detail, Figure \ref{fig:facc} shows that for $\beta \lesssim 1$ the solution boundary lies at optical depths a factor $\approx 10$ lower.  This prefactor remains order unity across (at least) the six orders-of magnitude in $\beta$ considered.

In physical units, this optically thin condition for solutions becomes $L_\infty \gtrsim  GM\dot{M}/r_B$.  Thus isothermal solutions must radiate at least the gravitational energy released on the way to the sonic point.  Since we expect accretion flows to release significantly more potential energy (down to the accretor's surface) we don't expect astrophysical solutions near this luminosity limit.

For the optically thick case, neighboring solutions are no longer isothermal.  We instead consider a rough requirement for solutions that $\partial_{\tilde{r}}\tilde{L} < 0$ in the outer hydrostatic regions where negative luminosities are triggered. The hydrostatic version of Eq.\ (\ref{eq:etot}) then requires $-\partial_{\tilde{r}}\tilde{T} \gtrsim  1/\tilde{r}^2$, again to order of magnitude.  For optically thick diffusion in outer regions where $\tilde{T} \sim \tilde{\rho} \sim 1$,  Eq.\ (\ref{eq:reqT}) gives $-d\tilde{T}/d\tilde{r} \sim \tilde{L}_\infty \tau/\tilde{r}^2$.  Negative luminosities are thus avoided down to $\tilde{r} \sim 1$ if $\tilde{L}_\infty \tau_B \gtrsim 1$.  This condition agrees with the solution boundary found numerically.

These estimates of valid solutions are based on the luminosity remaining positive at least to the sonic point. While negative luminosities developing interior to the sonic point (and thus underestimation of the extent of invalid models) is of potential concern, the discussion in \S \ref{sec:freefall} suggests this is largely avoided. When the flow is near freefall (as in the interior), $L$ is nearly constant (and thus non-negative) up to $r_{\rm src}$. This makes the luminosity most prone to becoming negative near the sonic point -- precisely the condition being applied above to characterize the `no-solution' regime.

Astrophysical solutions avoid this no-solution boundary as well.  At fixed $\tau_B$, luminosities on the optically thick boundary, $\tilde{L}_\infty \sim 1/ \tau_B$ are smaller by a factor $1/(\beta \tau_B^2) < 1$ (an inequality defined by the intersection of the boundaries) than the extension of the optically thin boundary.  As noted above, this higher luminosity already radiated insufficient gravitational energy.  We will show below (\ref{sec:thick}) that the neighboring optically thick solutions with low luminosity are not astrophysically relevant.

In summary, when $\tilde{L}_\infty$ is treated as a free parameter, regions of parameter space have no accretion solutions.  Astrophysical accretion flows with expected accretion luminosities naturally avoid this region of parameter space.  

\subsubsection{Isothermal Regime}\label{sec:iso}
Figure \ref{fig:facc} shows that rapid isothermal accretion with $f_{\rm acc} \approx 3$ occurs for $\tilde{L}_\infty\lesssim\min(1, 1/\tau_B)$, and $\tilde{L}_\infty \gtrsim \tau_B \beta$, the relevant branch of the allowed solution boundary (described above).  

These conditions require rapid cooling with $\beta \lesssim \min(1/\tau_B, 1/\tau_B^2)$.  The $\beta = 10^3$ panel in Figure \ref{fig:facc} shows no isothermal solutions, since the extension of the allowed ($\tilde{L}_\infty \gtrsim \tau_B \beta$) solutions to $\tilde{L}_\infty < 1$ requires $\tau_B \lesssim 10^{-3}$, outside the domain.   

The luminosity limit for isothermal accretion, $\tilde{L}_\infty \lesssim\ \min(1, 1/\tau_B)$, arises because high luminosities increase the temperature near the Bondi radius. For optically thick flows, isothermal conditions with $\tilde{T} \sim \tilde{\rho} \sim 1$ near 
$\tilde{r} \sim 1 $ require, from Eq.\ (\ref{eq:reqT}), 
$| \partial_{\tilde{r}} \tilde{T}| \sim \tilde{L}_\infty \tau_B \lesssim 1$.   
This criterion reproduces the isothermal optically thick boundary in the leftmost panel of Figure \ref{fig:facc}.  

Similarly, the optically thin boundary is explained by Eq.\ (\ref{eq:reqT}) with $\tau_B \lesssim 1$, giving isothermal conditions at  $\tilde{r} \sim 1$ for
$| \partial_{\tilde{r}} \tilde{T}| \sim \tilde{L}_\infty \lesssim 1$. 

These estimates assume both radiative equilibrium and constant $L$. From Eqs.\ (\ref{eq:radeq_analysis}), (\ref{eq:reqL}) with $f_{\rm acc} \sim 1$, these assumptions  require $\beta \lesssim 1$ and $\beta \tau_B \lesssim 1$, respectively.   While $\beta \tau_B \lesssim 1$ assures constant $L$ in this isothermal parameter regime (as shown above), radiative equilibrium may break down if $\beta \gtrsim 1$, but $\tau_B \lesssim 1/\beta \lesssim 1$.  We defer a more detailed analysis of this slow cooling regime, which is included in the numerical solutions.

We just showed that constant $L$ holds even without applying the expected accretion luminosities, as in \S \ref{sec:freefall}.  When we do consider the accretion luminosity, $\tilde{L}_\infty \sim \beta \tau/\tilde{r}_s$ from Eq.\ (\ref{eq:Linfacc}), with the accretion condition $\tilde{r}_s \lesssim 1$, we get  $\tilde{L}_\infty \gtrsim \beta \tau_B$.  This condition again reproduces the boundary between isothermal solution and the no-solution space.  Therefore, we do expect isothermal accretion solutions astrophysically, and they do not need to be isothermal interior to the sonic point.

\begin{figure*}
    \centering
    \includegraphics[width=\linewidth]{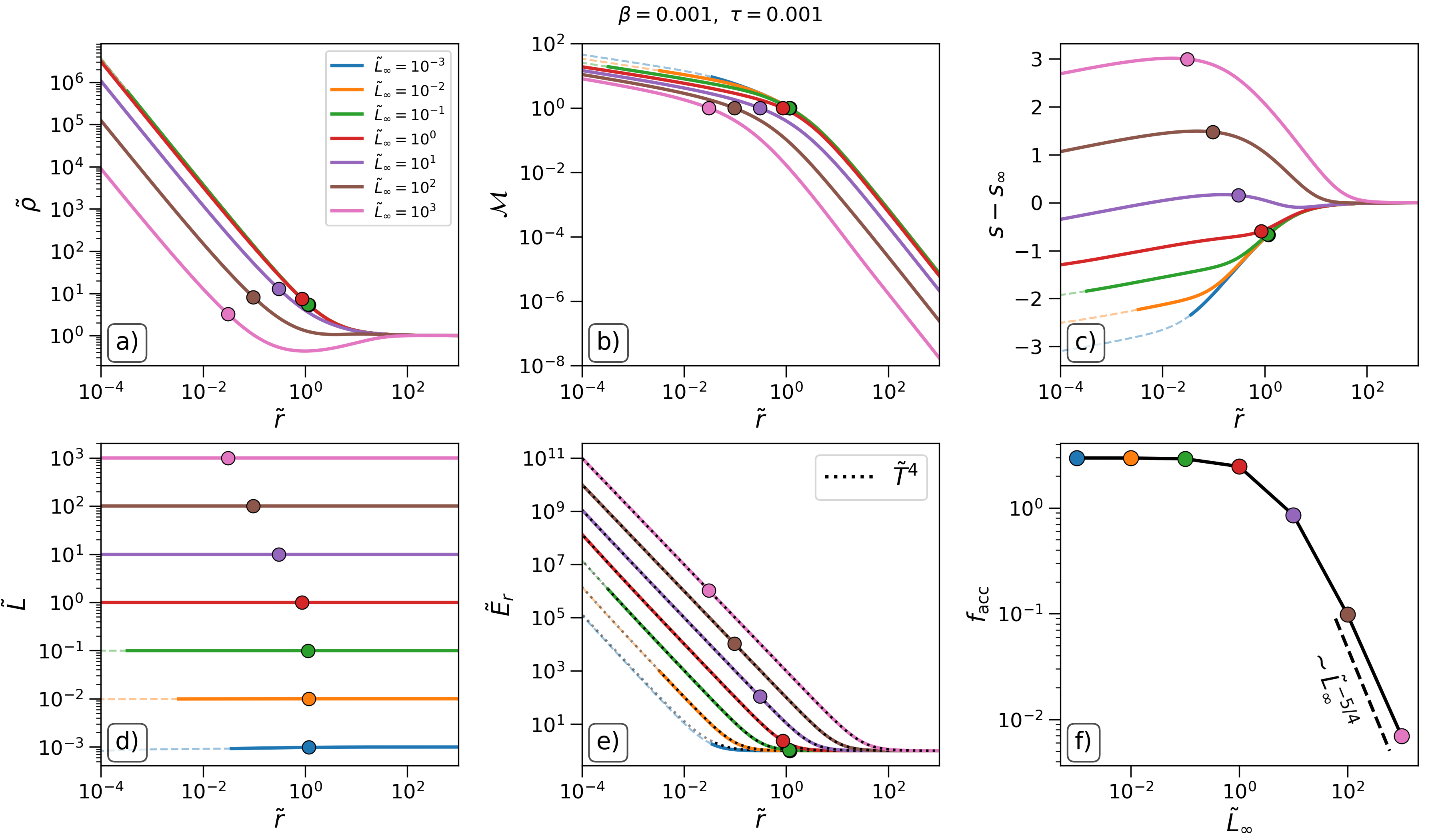}
    \caption{Panels (a-e): Radiating steady-state solutions for the models with $\tau_B=10^{-3}$, $\beta=10^{-3}$. Each panel plots the radial profile of a dependent variable of the problem $(\tilde{\rho}, \mathcal{M}, s, \tilde{L}, \tilde{E}_r)$ as solid lines with each curve corresponding to a different choice of luminosity $\tilde{L}_\infty$. Markers are placed at the sonic point of each curve. Along each curve we compute the physical radius of an accretor with self-consistent accretion luminosity according to Eq. (\ref{eq:Linfacc}). Inside this radius, the solution becomes unphysical and curves are drawn as faint dashed lines.} Panel (e) contains additional dotted lines for the temperature solutions $\tilde{T}^4$.
    Panel (f): The mass accretion rate normalized to the adiabatic rate $f_{\rm acc}$ as a function of model luminosity $\tilde{L}_\infty$.
    \label{fig:thin}
\end{figure*}

Figure \ref{fig:thin} shows the details of optically thin, rapidly cooling accretion solutions with $\tau_B=10^{-3}$, $\beta=10^{-3}$, which give isothermal accretion for $\tilde{L}_\infty \lesssim 1$.   These solutions are in radiative equilibrium (which was not enforced) with constant temperature to the sonic point.  The marginal $\tilde{L}_\infty = 1$ case is nearly isothermal with a modest decrease in $f_{\rm acc}$ and slight temperature increase at $r_B$.  The luminosity is effectively constant in all cases, which is expected since strong luminosity gradients develop near the valid solution boundary at $\tilde{L}_\infty \simeq \tau_B \beta = 10^{-6}$, well below the plotted values.  The largest entropy gradients are near the Bondi radius, with $ds/dr > 0$, indicating radiative cooling.  The non-isothermal $\tilde{L}_\infty \gtrsim 1$ solutions with suppressed accretion are addressed next.

\subsubsection{Optically Thin Regime ($\tau_B \lesssim \tilde{L}_\infty$)}\label{sec:thin}
We now explain the radiative suppression of Bondi accretion in the optically thin regime, using physical arguments and dimensional analysis. Since $\tilde{L}_\infty \lesssim 1$ gives isothermal solutions, the relevant parameter space in our overall Bondi problem  is $\tilde{L}_\infty \gtrsim 1$ and  $\tau_B \lesssim \tilde{L}_\infty$, as explained below. The simplified model in this section makes the radiative equilibrium and constant luminosity assumptions.  We check their validity afterwards.

Note also that while we refer to this regime as optically thin, the flow is idealized as infinite in extent and constant opacity, becoming optically thick $\rho\kappa r \propto r\gg 1$ as $r\rightarrow \infty$. At the same time, the deep interior has the flow near freefall, so $\rho\sim r^{-3/2}$ and $\rho\kappa r\sim r^{-1/2}$ -- again becoming optically thick. For $\tau_B \ll 1$ then, one has a traditional photosphere in the interior flow at $\tilde{r}\approx \tau_B^2$, followed by free-streaming at the scale of the Bondi radius until the flow becomes formally thick again at $\tilde{r}\approx 1/\tau_B$. For $\tau_B\gg 1$, the flow has $\rho\kappa r \gg 1 $ everywhere.

In this regime, the optically thin temperature is given by equation (\ref{eq:reqT}) with $C_1 = 1$ (constant $L$) and $\tau_B \rightarrow 0$, so that $\tilde{T}^4 = \tilde{L}_\infty/\tilde{r}^2$ in regions heated to $\tilde{T} \gtrsim 1$ .  Dimensionally this temperature  law gives, for $T \gtrsim T_\infty$ the standard
\begin{align}\label{eq:Tthin}
T &= \left(\frac{L_\infty}{16 \pi \sigma r^2} \right) ^{1/4} \equiv T_L \left(\frac{r_L}{r} \right)^{1/2}
\end{align} 
with Stefan-Boltzmann constant $\sigma$. Balancing gravitational and thermal energies at $r_L$ as $GM/r_L = \mathcal{R}T_L$, with ideal gas constant $\mathcal{R}$,  defines the temperature and length scales 
\begin{subequations}\label{eq:Trthin}
     \begin{align}
T_L &= \sqrt{\frac{L_\infty}{16\pi\sigma}} \frac{\mathcal{R}}{GM}=  T_\infty\left(\frac{\tilde{L}_\infty^{1/2}}{2\gamma}\right)
\\
r_L &= \left(\frac{GM}{\mathcal{R}  } \right)^2 \sqrt{\frac{16\pi\sigma}{L_\infty}} = \frac{GM}{\mathcal{R} T_L } = r_B \left(\frac{4 \gamma^2}{\tilde{L}_\infty^{1/2}}\right) \, .
\end{align} 
\end{subequations}
For $\tilde{L}_\infty \gtrsim 1$, these ``luminous" scales should approximately describe the temperature and location of the sonic point.  In scaled units, $\tilde{T}_L \sim 1/\tilde{r}_L \sim \tilde{L}_\infty^{1/2}$ to order unity.

Fig.\ \ref{fig:thin} confirms these scalings for the sonic point location, $\tilde{r} \simeq \tilde{r}_L\simeq \tilde{L}_\infty^{-1/2}$, and for the temperature and radiation energy density, $\tilde{E}_r \simeq \tilde{T}^4 \simeq \tilde{T}_L^4 \simeq \tilde{L}_\infty^2$ in $\tilde{L}_\infty \gtrsim 1$ models.

To explain the suppressed accretion rates, we also require a density scale.  Pressure equilibrium with the surroundings holds for $r \gtrsim r_L$, where  gravitational compression is weaker , as examined more below.  The ambient pressure,  $P_\infty$, thus sets the relevant density scale as $\rho_L \sim P_\infty/(\mathcal{R} T_L)$, or $\tilde{\rho}_L \sim 1/\tilde{T}_L$ when scaled. 
The sonic accretion rate is thus $\dot{M}\sim \rho_L c_L r_L^2$ with sonic speed $c_L\sim\sqrt{\mathcal{R}T_L}$. The resulting mass accretion rate,
 \begin{align}\label{eq:Mdot-thin}
\dot{M} &\simeq \frac{P_\infty (GM)^{9/2}}{\mathcal{R}^5 }\left(\frac{16 \pi \sigma}{L_\infty}\right)^{5/4}    \simeq \frac{\dot{M}_{\rm ad}}{\tilde{L}_\infty^{5/4}}
\end{align} 
with $f_{\rm acc} \simeq \tilde{L}_\infty^{-5/4}$, agrees with the suppressed accretion shown in Fig.\ \ref{fig:thin} for $\tilde{L}_\infty \gtrsim 1$.

We can also explain the density minima, which is most prominent in the $\tilde{L}_\infty = 10^3$ case.  At large $r$, the  $\mathcal{M} \ll 1$  hydrostatic limit with $T \propto r^{-1/2}$  gives:
\begin{align}
 \frac{d \ln \rho}{d \ln r} &=\frac{1}{2} - \frac{GM}{r \mathcal{R} T}    = \frac{1}{2} - \sqrt{\frac{r_L}{r} }  
\end{align} 
with a density minima at $r = 4 r_L = 16\gamma^2\tilde{L}_\infty^{-1/2} \simeq 0.99 (\tilde{L}_\infty/1000)^{-1/2}$.  This location again agrees with Fig.\ \ref{fig:thin}.  Inside this radius gravity is strong enough for the density to increase. Smaller $\tilde{L}_\infty$ values do not have strong density minima because the heated  $T \gtrsim T_\infty$ region does not extend to large enough radii. 

At high luminosities, solutions have a negative entropy gradient outside the sonic point. Thus the incoming subsonic flow is heated by the radiation. The entropy profile when $T\propto r^{-1/2}$ is given by pressure equilibrium with $d \ln \tilde{\rho}/d \ln \tilde{r} = 1/2$,  giving $ds/d \ln \tilde{r} = -7/10$ (since $\gamma = 7/5$).  We show in \S \ref{sec:convect} why these entropy gradients do not robustly lead to convection.

We also roughly predict the entropy value at the sonic point, With scaled $\tilde{T}_L \sim 1/\tilde{\rho}_L \simeq \tilde{L}_\infty^{1/2}$,  $s_L \sim (\gamma/2) \ln \tilde{L}_\infty$.  Fig.\ \ref{fig:thin} shows the expected entropy slope and the factor $0.7 \ln(10) \simeq 1.6$ jump in sonic point entropy for a factor 10 increase in $\tilde{L}_\infty$ (for $\tilde{L}_\infty \gtrsim 1$ models). In the interior, where the flow is near freefall we have $\tilde{\rho}\sim \tilde{r}^{-3/2}$, reversing the slope of the entropy profile $ds/d\ln \tilde{r}=1/10$. This corresponds to a $\Delta s=0.23$ per decade in $\tilde{r}$ matching the slopes in Fig.\ \ref{fig:thin}.

We now consider when the approximations used for these optically thin solutions hold, starting with optical depth itself.
The flow is optically thin near the sonic point if $\tilde{\rho}_L \tilde{r}_L \tau_B \lesssim 1$.  This condition, $\tau_B \lesssim \tilde{L}_\infty$, defines the boundary with the optically thick regimes discussed next.  Solutions become more optically thick away from the sonic point, with the local effective optical depth $\tilde{\rho} \tilde{r} \tau_B \propto \tilde{\rho} \tilde{r}$ increasing in both directions.  Thus solutions near this $\tau_B \sim \tilde{L}_\infty$ boundary will only be optically thin near the sonic point, leading to a gradual transition between optically thick and and thin behavior.

The radiative equilibrium approximation holds when the right-hand side of Eq.\ (\ref{eq:radeq_analysis}) is small. Thus radiative equilibrium holds near the sonic point (where it is most likely to break down) for $\beta \lesssim \tilde{r}_L/f_{\rm acc} \sim \tilde{L}_\infty^{3/4}$. We thus expect deviations from radiative equilibrium for $\tilde{L}_\infty \lesssim \beta^{4/3}$, roughly consistent with the numerical analysis of \S\ref{sec:radeq}. 

The assumption of constant $L$ exterior to the sonic point breaks down for $\tilde{L}_\infty \lesssim f_{\rm acc} \beta \tau_B \tilde{T}_L \sim (\beta \tau_B)^{4/7}$.   Combining with the optical depth criterion $\tau_B \lesssim \tilde{L}_\infty$   gives $\tau_B \lesssim \beta^{4/3}$.  Thus $L$ would vary only for $\tilde{L}_\infty \lesssim (\beta \tau_B)^{4/7} \lesssim \beta^{4/3}$, i.e.\ for conditions would also violate radiative equilibrium.  Thus we expect constant $L$ for all optically thin solutions in radiative equilibrium.  

Finally we apply the analysis of inner accretion solutions, from \S \ref{sec:freefall}, to this regime.  Combining the condition for an accretion solution with constant $L$ of  $\tilde{r}_s\lesssim \tilde{r}_\mathcal{M}\sim \tilde{r}_L$ with the consistent accretion luminosity $\tilde{L}_\infty \sim \beta \tau_B f_{\rm acc}/\tilde{r}_s$ gives the condition $\tilde{L}_\infty \gtrsim \beta \tau_B f_{\rm acc}/\tilde{r}_L \sim (\beta \tau_B)^{4/7}$, reproducing the constant $L$ which we just argued should hold in radiative equilibrium.

Eliminating $\tilde{L}_\infty$, again with the accretion solution, gives the condition  $\tilde{r}_s \lesssim (\beta\tau_B)^{-2/7}$ for an accretion solution.  The condition to be in this optically thin suppressed regime, $\tilde{L}_\infty \gtrsim \max(1, \tau_B)$, can similarly be expressed as $\tilde{r}_s \lesssim \min(\beta \tau_B, \beta/\tau_B^{5/4})$.  These conditions on $\tilde{r}_s$ can be checked for a given accretor size and mass in a specified background (disk model or otherwise).  Meeting them all is at least somewhat stricter than the  $\tilde{r}_s \lesssim 1$ condition to be inside the usual Bondi radius.

\subsubsection{Optically Thick Regime ($\tau_B \gtrsim\tilde{L}_\infty$)}\label{sec:thick}
We now seek to understand the radiative suppression of Bondi accretion in the optically thick regime.  As noted in \S\ref{sec:num} and shown in  Figure \ref{fig:facc}, optically thick accretion shows two distinct behaviors:  (1) diagonal accretion rate contours at higher luminosities, i.e.\ dependent on both $\tilde{L}_\infty$ and $\tau_B$ and (2) horizontal $f_{\rm acc}$ contours at lower luminosities, which depend on $\tau_B$ and also $\beta$.  

\begin{figure*}
    \centering
    \includegraphics[width=\linewidth]{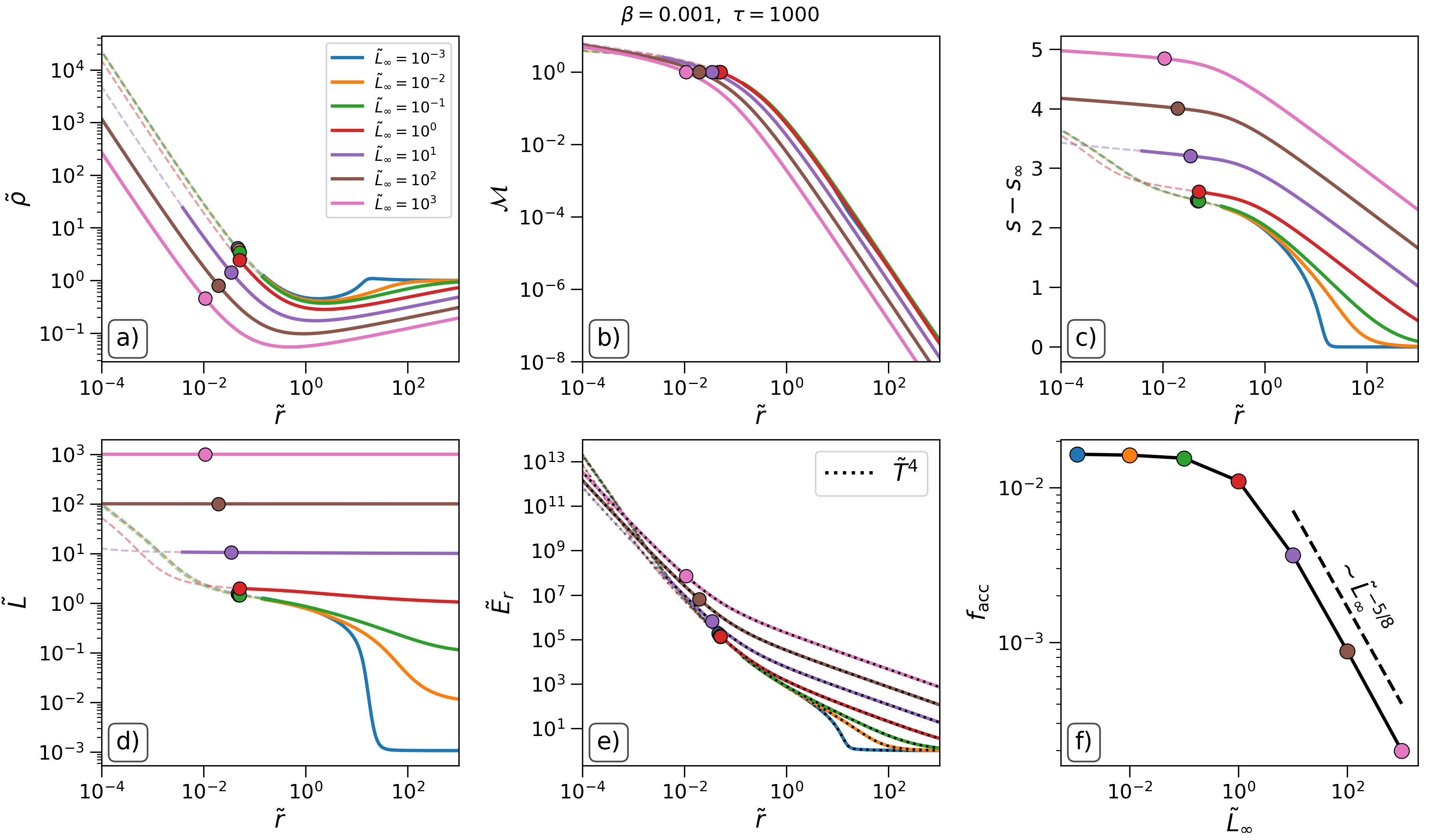}
    \caption{Same   as Figure \ref{fig:thin} but for models with $\tau_B=10^3$, $\beta=10^{-3}$}
    \label{fig:thick}
\end{figure*}

Figure \ref{fig:thick} shows detailed accretion solutions for a range of luminosities in the optically thick and radiative equilibrium regime ($\tau_B=10^3$, $\beta=10^{-3}$).  For the higher luminosities the accretion rate follows $f_{\rm acc}\sim (\tilde{L}_\infty\tau_B)^{-5/8}$,  while lower luminosities show $f_{\rm acc} \sim (\beta \tau_B^2)^{-5/11}$ independent of $\tilde{L}_\infty$.

To explain these behaviors, and derive these scalings, we use approximate radiative equilibrium scalings, as above. Figure \ref{fig:thick} shows the larger luminosity cases have constant $\tilde{L} \simeq \tilde{L}_\infty$, while the lower $\tilde{L}_\infty$ cases have varying luminosity, with $\tilde{L} \gg \tilde{L}_\infty$ near the sonic radius.  We use this insight to develop the analytic models for these two cases.   We will address the parameter space boundaries of these optically thick cases in \S\ref{sec:unify} and App.\ \ref{app:formula}.

\paragraph{Case 1: High Luminosity}
Similar to the optically thin case, we consider the relevant temperature law for heated ($\tilde{T} \gtrsim 1$) regions in pressure equilibrium ($\tilde{\rho} \simeq 1/\tilde{T}$ and constant luminosity.  Equation (\ref{eq:reqT}), now with $C_1\rightarrow 0$ for large $\tau_B$, gives 
\begin{align}\label{eq:Trho_thickouter}
\tilde{T} &\simeq  \frac{1}{\tilde{\rho}} \simeq \left(\frac{15 \tilde{L}_\infty\tau_B}{4 \tilde{r}}\right)^{1/5} = \frac{T_L}{T_\infty} \left(\frac{r_L}{r} \right)^{1/5}  
\end{align}
which matches temperature power-laws seen in the plots of $\tilde{E}_r = \tilde{T}^4$ in Figure \ref{fig:thick}, for higher $\tilde{L}_\infty$. The relevant temperature and density scales are again set by   $GM/r_L = \mathcal{R}T_L$, which, in this case, gives
\begin{subequations}
    \begin{align}
T_L &= \left(\frac{15 L_\infty P_\infty \kappa}{64\pi G M \sigma }\right)^{\frac 1 4} =  {T_\infty}\left(\frac{15 \tilde{L}_\infty \tau_B }{4(2\gamma)}\right)^{\frac 1 4}
\\
r_L &= \frac{1}{\mathcal{R}  } \left(\frac{64\pi\sigma(GM)^{5}}{15 L_\infty P_\infty \kappa}\right)^{\frac 1 4}  = r_B \left(\frac{4(2\gamma)^5}{15 \tilde{L}_\infty \tau_B }\right)^{\frac 1 4} .
\end{align} \end{subequations}
These temperature and radius scales describe how, for higher luminosities, the sonic point moves radially inward, and gets hotter, as seen in Figure \ref{fig:thick}. This luminosity dependence, $\tilde{L}_\infty^{\pm 1/4}$ respectively, is weaker than for the optically thin case, Eq.\ (\ref{eq:Trthin}).  The temperature profile and resulting scales would change for different opacity laws; this constant opacity case is a simple example.

The characteristic density, $\rho_L = P_\infty/(\mathcal{R} T_L)$ depends on the ambient pressure as
\begin{align}
\rho_L & =  \left(\frac{GM \sigma}{L_\infty \kappa } \right)^{1/4} \frac{P_\infty^{3/4}}{\mathcal{R}} = \rho_\infty \left(\frac{2\gamma}{\tilde{L}_\infty\tau_B}\right)^{1/4}\, .
\end{align}
 This density scale explains how the density minima (where gravity becomes significant, near $r_L$) and the sonic point densities (larger, but closely related) both decrease with $\tilde{L}_\infty$, as seen in Figure \ref{fig:thick}.  

The characteristic values of the dimensionless variables, to order unity, are thus $\tilde{T}_L \sim 1/\tilde{\rho}_L \sim 1/\tilde{r}_L \sim (\tilde{L}_\infty \tau_B)^{1/4}$, which is useful for estimates.  For example, these solutions are optically thick for $\tau_B \tilde{\rho}_L \tilde{r}_L \gtrsim 1$ or $\tau_B \gtrsim \tilde{L}_\infty$, consistent with the optical depth boundary found from the optically thin solutions.

These scales give the approximate accretion rate, $\dot{M}\simeq  \rho_L \sqrt{\mathcal{R}T_L} r_L^2$, as
\begin{equation}
\begin{split}
\dot{M}  &\simeq \frac{(GM)^{ {21} /8} P_\infty ^{3/8}}{ \mathcal{R}^{{5}/{2} } } \left(\frac{L_\infty \kappa }{\sigma}\right)^{-5/8}\\ 
&\sim \frac{\dot{M}_{\rm ad}}{(\tilde{L}_\infty \tau_B)^{5/8}}
\end{split}
\end{equation}
This result confirms $f_{\rm acc} \simeq (\tilde{L}_\infty \tau_B)^{-5/8}$, shown in Fig.\ \ref{fig:thick}.  
In physical units, the dependence of accretion rate on ambient pressure (and thus density) is weaker than the linear dependence of both standard Bondi accretion and optically thin suppression.  Physically, this effect occurs because higher pressure also lead to higher temperatures and smaller accretion radii, as $T_L \propto 1/r_L \propto P_\infty^{1/4}$.  The luminosity scaling is also weaker compared to the optically thin case -- see Eq.\ (\ref{eq:Mdot-thin}).

These optically thick models also have negative entropy gradients outside the sonic point, where the power-law slope in the outer heated regions is $ds/d \ln \tilde{r} = \gamma d \ln T/d\ln r = -7/25$.  The sonic point entropy is roughly
\begin{align}
s &\simeq  \gamma \ln\tilde{T}_L \simeq \frac{7}{20} \ln(\tilde{L}_\infty\tau_B)  
\end{align}
Thus for every decade increase in $\tilde{L}_\infty$ (or $\tau_B$), we expect $s$ at the sonic point to increase by $0.35 \ln(10) \simeq 0.8$ in good agreement with Fig.\ \ref{fig:thick}.

For inner regions in freefall with $\rho \propto r^{-3/2}$, constant opacity and constant $L$ radiative diffusion transitions to a steeper $T \propto r^{-5/8}$.    Free fall regions would thus have a flat entropy gradient  $ds/d\ln \tilde{r} \rightarrow (\gamma -1)(3/2) - 5/8 = -1/40$.
This temperature steepening and entropy flattening is already seen approaching the sonic point in Fig.\ \ref{fig:thick}.

Applying these scalings to Eq.\ (\ref{eq:radeq_analysis}) shows that radiative equilibrium holds for $\beta \lesssim \tilde{r}_L/f_{\rm acc} \simeq (\tilde{L}_\infty \tau_B)^{3/8}$. Constant $L$ holds for $\tilde{L}_\infty \gtrsim f_{\rm acc} \beta \tau_B \tilde{T}_L \sim \beta^{8/11} \tau_B^{5/11}$. This constant $L$ condition is stricter since it can be rewritten as $\beta < (\tilde{L}_\infty\tau_B)^{3/8} (\tilde{L}_\infty/\tau_B)$, and $\tilde{L}_\infty/\tau_B \lesssim 1$ for these  optically thick solutions.  In conclusion the parameter space of high-luminosity optically thick solutions is $\beta^{8/11}\tau_B^{5/11}\lesssim\tilde{L}_\infty\lesssim\tau_B$, where furthermore the radiative equilibrium approximation is good.

Finally we apply the analysis of inner accretion solutions, from \S \ref{sec:freefall}, to this regime.  As in the optically thin case, a valid accretion solution must have $\tilde{L}_\infty \gtrsim \beta \tau_B f_{\rm acc}/\tilde{r}_L$. With the derived scalings for $r_L$, $f_{\rm acc}$, this condition becomes $\tilde{L}_\infty \gtrsim \beta^{8/11}\tau_B^{5/11}$ -- the same as the condition for constant luminosity. We conclude that solutions with non-constant luminosity are not valid accretion solutions in the sense that they do not admit a self-consistent interior solution. 
Using the accretion luminosity, Eq.\ (\ref{eq:Linfacc}), the condition for a solution in terms of accretor size is $\tilde{r}_s\lesssim(\beta \tau_B^2)^{-2/11}$.  Consistent with the analysis of optically thin solutions, the solutions are only optically thick for $\tilde{r}_s \gtrsim \beta/\tau_B^{5/4}$.

\paragraph{Case 2: Low, Radially Varying Luminosity} 
In the preceding case, it was seen that cases with substantially varying luminosity do not admit valid accretion solutions. At the same time, by treating $\tilde{L}_\infty$ as a free parameter, we do recover \textit{exterior} solutions with radially varying luminosity across the computed parameter space (Figure \ref{fig:thick}). For the sake of completeness, we treat these non-constant luminosity regimes here but note that astrophysical applications are limited, since they do not admit viable interior solutions. 

We apply Eq.\ (\ref{eq:reqL}) to the case of pressure equilibrium  and heated regions with $\tilde{T} \gtrsim 1$ and $\tilde{L} \gtrsim \tilde{L}_\infty$ so that  
\begin{align}\label{eq:Lcase2}
\tilde{L}\approx f_{\rm acc}'\beta\tau_B\tilde{T} /3
\end{align} 
where $ f_{\rm acc}' \equiv 12\gamma q_\gamma f_{\rm acc} $ hides order unity constants that we will ignore in this scaling analysis.

To find the temperature profile, we apply equation (\ref{eq:Lcase2})  to equation (\ref{eq:reqT}), with $C_1 = 0$,\footnote{This assumption drops a $dL/dr$ term, which can be justified after the fact.} giving 
\begin{equation}
    \tilde{T}\simeq \frac{1}{\tilde{\rho}}\simeq \left(
    \frac{ f_{\rm acc}'\beta\tau_B^2}{\tilde{r}}\right)^{1/4} =  \frac{T_L}{T_\infty} \left(\frac{r_L}{r} \right)^{1/4}     
\end{equation}
where the characteristic scales, given again by $\mathcal{R}T_L = GM/r_L $, are now
\begin{subequations}
    \begin{align}
T_L &= \left(\frac{f_{\rm acc}'\beta \tau_B^2}{2\gamma} \right)^{1/3} T_\infty \sim \left(\frac{\mathcal{R} \dot{M}\kappa P_\infty}{\sigma GM}\right)^{1/3} \\
{r}_L &= \frac{(2\gamma)^{4/3}}{(f_{\rm acc}'\beta \tau_B^2)^{1/3}}r_B \sim  \left(\frac{\sigma (GM)^4}{\mathcal{R}^4 \dot{M}\kappa P_\infty } \right)^{1/3}
\end{align} \end{subequations}
where the dimensional values ignore order unity factors.  These scales depend on the accretion rate due to  steady state energy balance, Eq.\ (\ref{eq:AYs}).  Figure \ref{fig:thick} shows that varying the escaping $\tilde{L}_\infty$, in this low  $\tilde{L}_\infty$ regime, does not affect the accretion solution, and it does not affect these scales.

To order unity the characteristic values of these scaled variables are $\tilde{T}_L \sim 1/\tilde{\rho}_L \sim 1/\tilde{r}_L \sim (f_{\rm acc} \beta \tau_B^2)^{1/3}$, again assuming pressure equilibrium.  These scales  still contain the accretion rate, which is found as
\begin{align}
f_{\rm acc} &\sim \tilde{\rho}_L \tilde{r}_L^2 {\tilde{T}_L}^{1/2} \sim (f_{\rm acc} \beta \tau_B^2)^{-5/6} \sim (\beta \tau_B^2)^{-5/11}\, .
\end{align} 
This accretion rate explain the trends in Figure \ref{fig:facc} in this regime.  Specifically the horizontal contours of $f_{\rm acc}$ are independent of $\tilde{L}_\infty$ with values that decrease with both $\beta$ and $\tau_B$.  The relevant estimate for Figure \ref{fig:thick} is $f_{\rm acc} \sim 10^{-3\cdot 5/11} \simeq 0.04$, which is order-of-magnitude consistent with the $f_{\rm acc} \simeq  0.015$ plateau at low luminosity.

In physical units, this accretion rate is
\begin{equation}
        \dot{M} \sim \left[ P_\infty (GM)^{17} \left(\frac{\sigma}{\mathcal{R}^4\kappa }\right)^5\right]^{1/11} 
\end{equation}
where the extremely weak dependence on ambient pressure emphasizes the thermal regulation.  We again emphasize that we do not expect these solutions to be astrophysically relevant, since accretion luminosities should be too large.  

\subsection{Summary of Accretion Rate Scalings}\label{sec:unify}
Having found appropriate scalings needed to describe the behavior of the radiative equilibrium models of Fig.\ \ref{fig:facc}, here we summarize the regimes and the associated scalings. This summary is intended to most accurately represent radiative equilibrium conditions, where $\beta$ values are sufficiently small as detailed above in Fig.\ \ref{fig:diseq} and with analytic scalings. Paper  II will show that radiative equilibrium is valid for our application to planetary accretion.

\begin{figure}
    \centering
    \includegraphics[width=\linewidth]{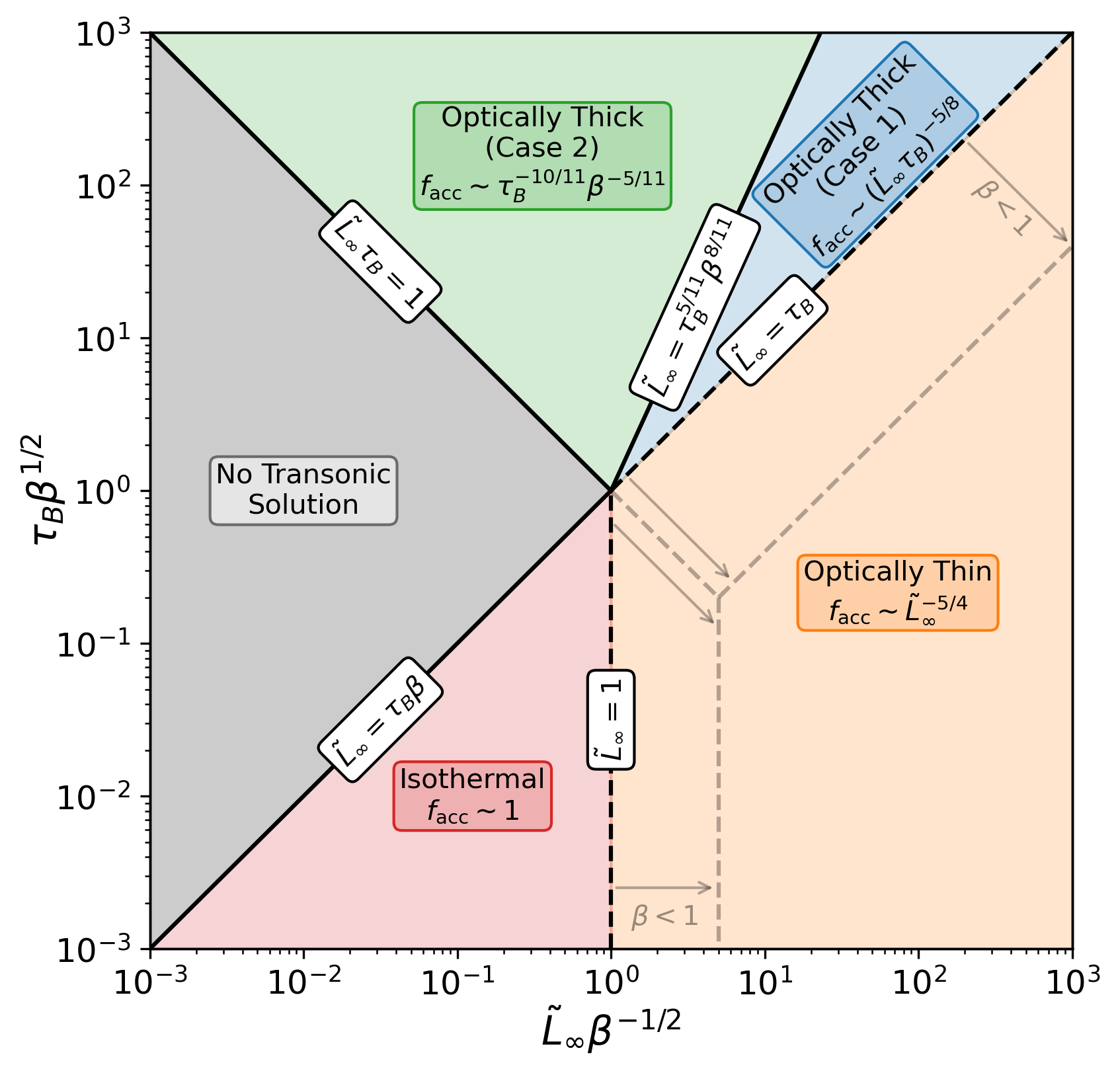}
    \caption{A summary of the approximate regimes and associated scalings for models satisfying radiative equilibrium. Lines mark the boundaries between the regimes described in \S\ref{sec:nosol}-\ref{sec:thick}. Solid lines are fixed in the scaled $(\tilde{L}_\infty\beta^{-1/2},\tau_B\beta^{1/2})$ coordinate system chosen here. Dashed lines are translated in these coordinates based on the value of $\beta$ -- the thick dashed line assumes $\beta=1$ with the faint lines showing the effect of $\beta<1$ in these scaled coordinates.}
    \label{fig:scalings}
\end{figure}

Using Figure \ref{fig:facc} and Section \ref{sec:nosol} as a starting point, we note that the different mass accretion regimes are somewhat reasonably divided by the contours for $\tilde{L}_\infty\tau_B=1$ and $\tilde{L}=\tau_B\beta$. In the preceding sections it was demonstrated that these contours separate the ``no-solution'' regime from neighboring optically thick and isothermal solutions. The intersection of these contours at point $(\tau_B,\tilde{L})=(\beta^{-1/2}, \beta^{-/2})$, thus provide a convenient fixed point from which to anchor the solutions. In Figure \ref{fig:scalings} we map the compiled scalings of the previous sections relative to this fixed point -- that is, scaled coordinates $(\tilde{L}_\infty\beta^{-1/2},\tau_B\beta^{1/2})$. In this coordinate system, the ``no-solution'' boundaries $\tilde{L}\tau_B =1$, $\tilde{L}=\tau_B\beta$, and the optically thick boundary $\tilde{L}_\infty =\tau_B^{5/11}\beta^{8/11}$ all remain fixed as $\beta$ is varied. Boundaries defining the optically thin regime ($\tilde{L}_\infty= 1$ and $\tilde{L}_\infty=\tau_B$) are not fixed in this coordinate system and shift as $\beta$ is varied. Figure \ref{fig:scalings} shows good qualitative agreement with Figure \ref{fig:facc} for regimes in radiative equilibrium ($\beta\lesssim 1$). In written form these scalings and regimes are summarized as:
\begin{equation*}
    f_{\rm acc} \sim 
    \begin{cases}
    \text{No solution}  & \tilde{L}_\infty/\beta\lesssim\tau_B\lesssim \tilde{L}_\infty^{-1}\\
    1 &\tau_B\beta\lesssim\tilde{L}_\infty\lesssim \min(1,\tau_B^{-1})\\
    \tilde{L}_\infty^{-5/4} &\tilde{L}_\infty\gtrsim \max(1, \tau_B) \\
    f_{\rm thick} & \tau_B^{-1} \lesssim \tilde{L}_\infty \lesssim \tau_B\\
    \end{cases}
\end{equation*}
\begin{equation*}
    f_{\rm thick} \sim 
    \begin{cases}
    (\tilde{L}_\infty\tau_B)^{-5/8} &\tilde{L}_\infty \gtrsim \tau_B^{5/11} \beta^{8/11} \\
    (\beta \tau_B^2)^{-5/11} &\tilde{L}_\infty\lesssim\tau_B^{5/11}\beta^{8/11} \\
    \end{cases}
\end{equation*}

For a more quantitative treatment, we refer the reader to Appendix \ref{app:formula} where we use the numerical solutions to better fit order unity normalizations in the above scalings. This produces an analytic formula for $f_{\rm acc}$ which more accurately fits the radiative equilibrium solutions.

\section{Applicability \& Caveats}\label{sec:app}

\subsection{Boundary Conditions}\label{sec:bc}
The numerical solutions of Section \ref{sec:num} were computed assuming an idealized Bondi scenario in which the outer boundary is effectively infinite. To make the problem numerically tractable but still heavily idealized, the outer boundary was placed at $r=10^{12}r_B$, far larger than any astrophysical extent of interest. Here we verify that this exceedingly large idealized outer boundary condition is not determining the results to any worthwhile extent. 

\begin{figure*}
    \centering
    \includegraphics[width=\linewidth]{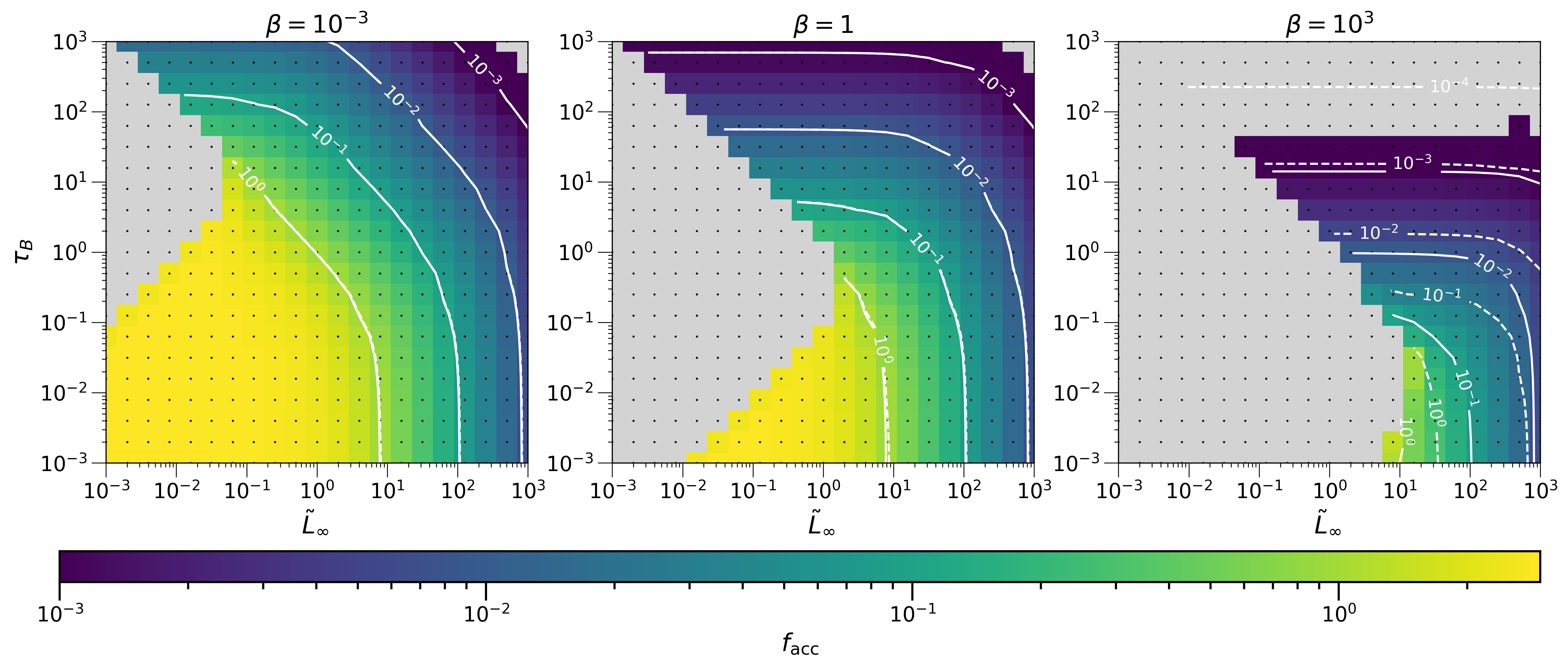}
    \caption{Maps of $f_{\rm acc}$ constructed in the style of Figure \ref{fig:facc}, but for models which place the outer boundary at $r=10^{2}r_B$. Solid contours correspond to $f_{\rm acc}$ for models with this small outer boundary, while dashed contours show the accretion rates of the fiducial large $r=10^{12}r_B$ outer boundary models (Fig.\ \ref{fig:facc}).}
    \label{fig:bc}
\end{figure*}

To do so, we compute numerical solutions over the parameter space of Section \ref{sec:num}, but we move the outer boundary inwards to a more reasonable $r=10^2r_B$. The boundary conditions for each model are kept the same, that is -- their values at infinity $(\rho, s,L,E_r)=(\rho_\infty, s_\infty, L_\infty, aT_\infty^4)$. Since we set the boundary at a modest $r=10^2r_B$, we solve only the radiative disequilibrium equations (\ref{eq:dim-rho})-(\ref{eq:dim-er}), rather than perform the switching procedure described in Appendix \ref{app:inteq}. The results of these integrations are plotted in Figure \ref{fig:bc} for comparison with the fiducial models in Figure \ref{fig:facc}. We find this na\"{i}ve boundary condition makes our solution procedure a bit less reliable, hence a few more models were unable to return a solution, particularly at large $\tau_B$, $\tilde{L}_\infty$. Nevertheless, the models with transonic solutions show very good agreement with the fiducial solutions. The only regime where noticeable differences are found are for $\beta\gg1$. In this regime, biased toward models with strong disequilibrium, there are noticeable but not discrepant differences in the computed accretion rates. These differences are still less than an order of magnitude, show the same qualitative radiation suppression behavior, and lessen as the models approach radiative equilibrium $\tau_B\rightarrow \infty$. Thus we conclude that the fiducial models and associated results may be applied to astrophysical systems without much concern for the outer extent of the flow. This also implies that instead of thinking of the properties which determine the accretion rates as some abstract ``properties at infinity", they are more reasonably thought of as characteristic properties -- density, opacity, etc. -- set by some large scale physics of the environment.

\subsection{Stability to Convection}\label{sec:convect}
As demonstrated by the accretion solutions in Figures \ref{fig:thin} \& \ref{fig:thick}, a reduction in the accretion rate is coincident with a negative entropy gradient, at least outside the sonic point. In static equilibrium, such as in stellar interiors, such a gradient corresponds to negative squared Brunt-V{\"a}is{\"a}l{\"a} frequency $N^2\equiv g(ds/dr)/\gamma$ (with $g\equiv GM/r^2>0$) and convective instability. This criterion for convection is equivalently framed as a requirement that the temperature gradient $\nabla_T \equiv \partial \ln T/\partial \ln P$ be steeper than the adiabatic temperature gradient under adiabaticity $\nabla_{\rm ad} \equiv (\partial \ln T/\partial \ln P)_{\rm ad} = (\gamma-1)/\gamma$, i.e.\ the Schwarzschild criterion $\nabla_T - \nabla_{\rm ad} > 0$. In our case, the Schwarzschild criterion is not altogether appropriate as the background state is accreting, not static, prompting us to consider a more dedicated convective analysis.

The propensity for Bondi flows to develop adverse entropy gradients was noted by \citet{Flammang1984} and predicted to be a generic possibility for gas pressure-dominated flows. It was predicted that interior to the sonic point, shock dissipation from supersonic motions would make any convective energy transport negligibly inefficient. Exterior to the sonic point, a necessary but not sufficient condition for convection to operate requires the characteristic growth time of convective instabilities $t_{\rm conv} \equiv 1/\sqrt{-N^2}$ be faster than the advective timescale with which fluid elements are vertically elongated and horizontally compressed by the background accretion flow $t_{\rm adv} \equiv r/v$. Simulations and stability analyses of core-collapse supernovae suggest the similar condition $t_{\rm adv}/t_{\rm conv} \gtrsim 3$ \citep{Foglizzo+2006}. 

At the same time, convection may also be inefficient and unlikely to override the background entropy profile. When the radiative cooling time of a convective blob is shorter than the characteristic convective timescale, we expect convection to be too inefficient to effectively perturb the background. The radiative cooling time for a perturbation with wavenumber $k$, specific heat capacity at constant pressure $c_P$, and Planck/Rosseland mean opacities $\kappa_P,\kappa_R$ is estimated as \citep{Spiegel1957, UnnoSpiegel1966, MihalasMihalas1984},
\begin{equation}
    t_{\rm cool}^{-1}(k) =  \frac{4a_rc\kappa_P T^3}{c_P}\left(\frac{1}{1 + 3\rho^2\kappa_R\kappa_P/k^2}\right).
\end{equation}
For a convective blob, the wavenumber ought to be scaled to an inverse pressure scale-height $k\sim 1/H_p$. Therefore, for convection to significantly effect the solutions of Section \ref{sec:models} requires both $t_{\rm conv} \lesssim  t_{\rm adv}$ and $t_{\rm conv}\lesssim t_{\rm cool}$. 

This timescale argument for convective instability is more precisely formulated in  \citet{Markovic1995}, where a mixing-length model and instability criterion are developed specifically for the problem of Bondi accretion. In this formalism, the radial stretching of convective fluid elements counteracts the buoyant action unless a sufficiently strong luminosity gradient (i.e. entropy gradient, equation \ref{eq:AYs}) also acts. The Schwarzschild criterion for instability $\nabla_T - \nabla_{\rm ad} > 0$, is thus modified to a stricter
\begin{equation}\label{eq:markcrit}
    \nabla_T - \nabla_{\rm ad} > 2 U\Psi
\end{equation}
where $U$ quantifies the heat transport of a convective blob and $\Psi$ contains most of the ``stretching" action. To apply this formalism to our context, we make one modification which is to extend the heat transport term $U$ to optically thin regimes. In the spirit of \citet{Spiegel1957, Henyey+1965, UnnoSpiegel1966}, $U$ is modified from \citet{Markovic1995} equation (31) to, 
\begin{equation}
U = \frac{1}{\bar{g}}\left(\frac{8P}{\rho \delta}\right)^{1/2}\left(\frac{ac\kappa_P T^3}{c_P}\right)\left(\frac{1}{1+\kappa_R\kappa_P\rho^2 l_m^2/3}\right)
\end{equation} 
with an ``effective" gravity $\bar{g}\equiv GM/r^2 + v(dv/dr)$, $l_m$ the mixing length, and $\delta\equiv -(\partial\ln \rho/\partial\ln T)_P = 1$ for an ideal gas. Meanwhile, $\Psi$ is kept as 
\begin{equation}
    \Psi = \frac{1}{2}\frac{l_m}{w}\frac{dv}{dr}
\end{equation}
where 
\begin{equation}
    w \equiv \left(\frac{\delta \bar{g} l_m^2}{8H_p}\right)^{1/2}
\end{equation}
with pressure scale height $H_P$. Ignoring factors of order unity and approximating $l_m\approx H_P$, $dv/dr\approx v/r \rightarrow t_{\rm adv}^{-1}$, $g\approx \bar{g}$, the criterion (\ref{eq:markcrit}) is of the form
\begin{equation}
    t_{\rm conv}^2 \lesssim \left(\frac{gH_P}{c_s^2}\right)^{1/2}t_{\rm cool} t_{\rm adv}
\end{equation}
demonstrating that $t_{\rm conv} \lesssim t_{\rm adv}, t_{\rm cool}$ is qualitatively appropriate, especially when $gH_P/c_s^2\approx 1$ for hydrostatic equilibrium. 

To estimate the potential importance of convection across our parameter space, we evaluate our modified Markovic criterion for the models of Section \ref{sec:num} taking $l_m=H_P$. In general, we find that models with significant accretion suppression are convectively unstable at sufficiently large radii, where weak flow corrections cannot overcome negative entropy gradients. At smaller radii however, advection and/or cooling times become sufficiently short to make the flow convectively stable. Thus most convectively unstable models have a radiative-convective boundary (RCB) at some radius, $r_{\rm RCB}$, inside which the flow is convectively stable, at least to the sonic point.

\begin{figure}
    \centering
    \includegraphics[width=\linewidth]{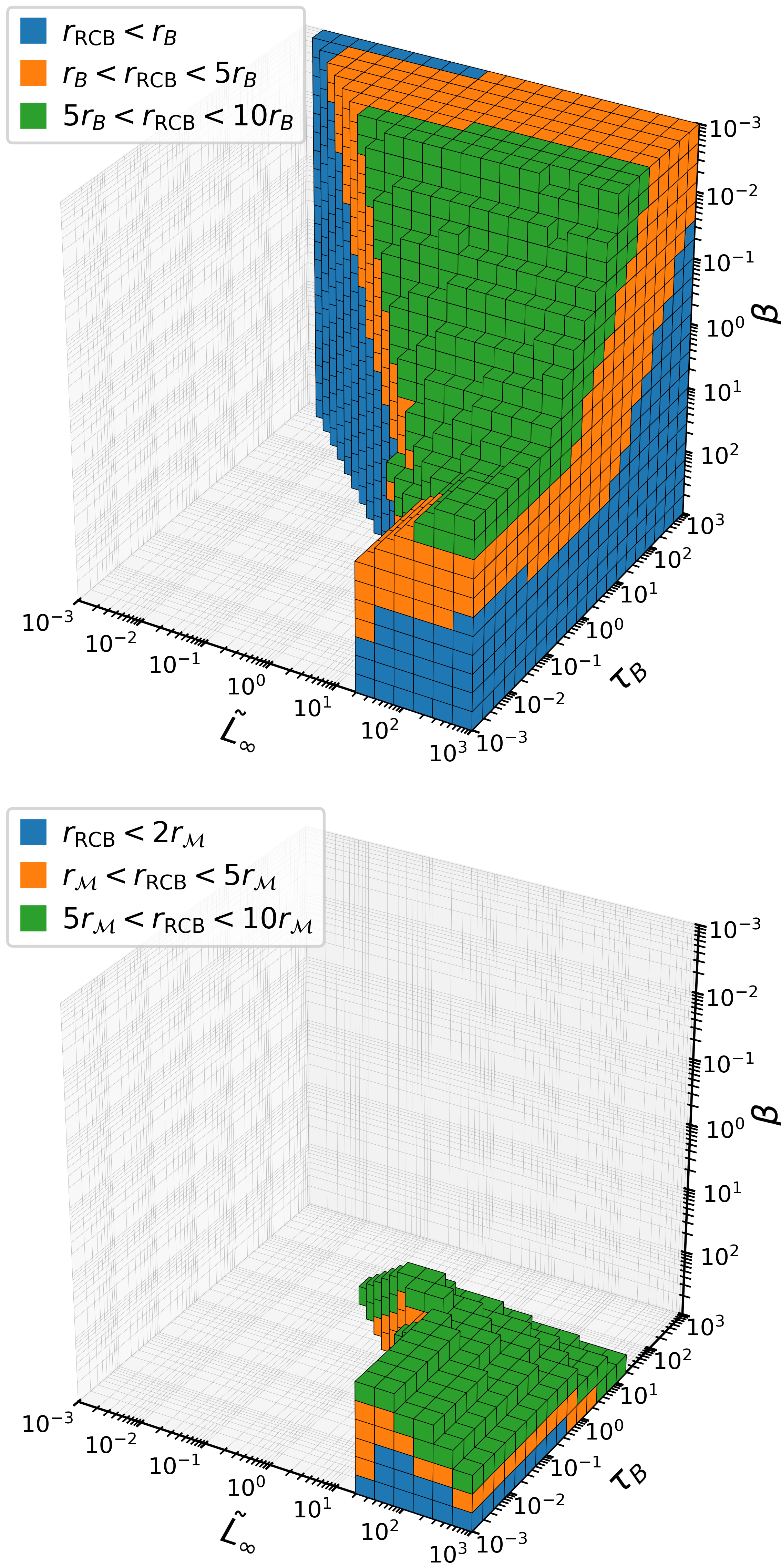}
    \caption{The approximate scale of the computed radiative-convective boundary across our parameter space. Shaded models in the upper panel are binned based on the scale of $r_{\rm RCB}$ relative to the nominal Bondi radius and all have the RCB occurring within the inner $10r_B$. In feedback suppressed cases, however, the sonic point is moved interior to the nominal Bondi radius. Therefore, in the lower panel, the same models are instead binned by RCB location relative to the sonic radius. The $\beta$-axis here is inverted as in Figure \ref{fig:diseq}.}
    \label{fig:conv3d}
\end{figure}

We plot the locations of the RCB in the radiative solutions of Section \ref{sec:num} in Figure \ref{fig:conv3d}. The upper panel indicates the location of RCBs that fall within 10 $r_B$. Not surprisingly, the models with the deepest outer convective zones (smallest $r_{\rm RCB}$) have long cooling times and/or large optical depths and also fall in the parameter space where accretion rates are suppressed.

However, because the sonic point is located well inside $r_B$ in models with suppressed accretion, the bottom panel of Figure \ref{fig:conv3d} shows which models have RCBs close to the computed sonic point (with $r_{\rm RCB} < 10r_{\mathcal{M}}$. Only models with long cooling times, i.e.\ large $\beta$, have RCBs close to the sonic point.

Since outer convective zones start far from the critical sonic point in most models, we expect these convective zones to have a small effect on accretion rates. The role of convection is best studied by direct numerical simulation or mixing-length models.  We defer such detailed models and instead constrain the impact of convection as follows.  

We consider a simple model of efficient convection, where convectively unstable regions adjust to adiabatic flow with $\partial_r s=0$.  While this approximation is known to be accurate in many stellar and planetary interiors, it overestimates the role of convection, by neglecting both inefficient convection and corrections of the Markovi\'{c} criterion. Both of these effects would give steeper entropy gradients, closer to our purely radiative solutions.

To apply this approximation, we integrate the adiabatic equations (\ref{eq:adiabatic_eqns}) in convectively unstable regions and transition to our radiative equations (\ref{eq:dimless}) inside the RCB.  In this piecewise system, we vary $\dot{M}$ to find the critical solution that passes through a sonic point, as usual.

\begin{figure*}
    \centering
    \includegraphics[width=\linewidth]{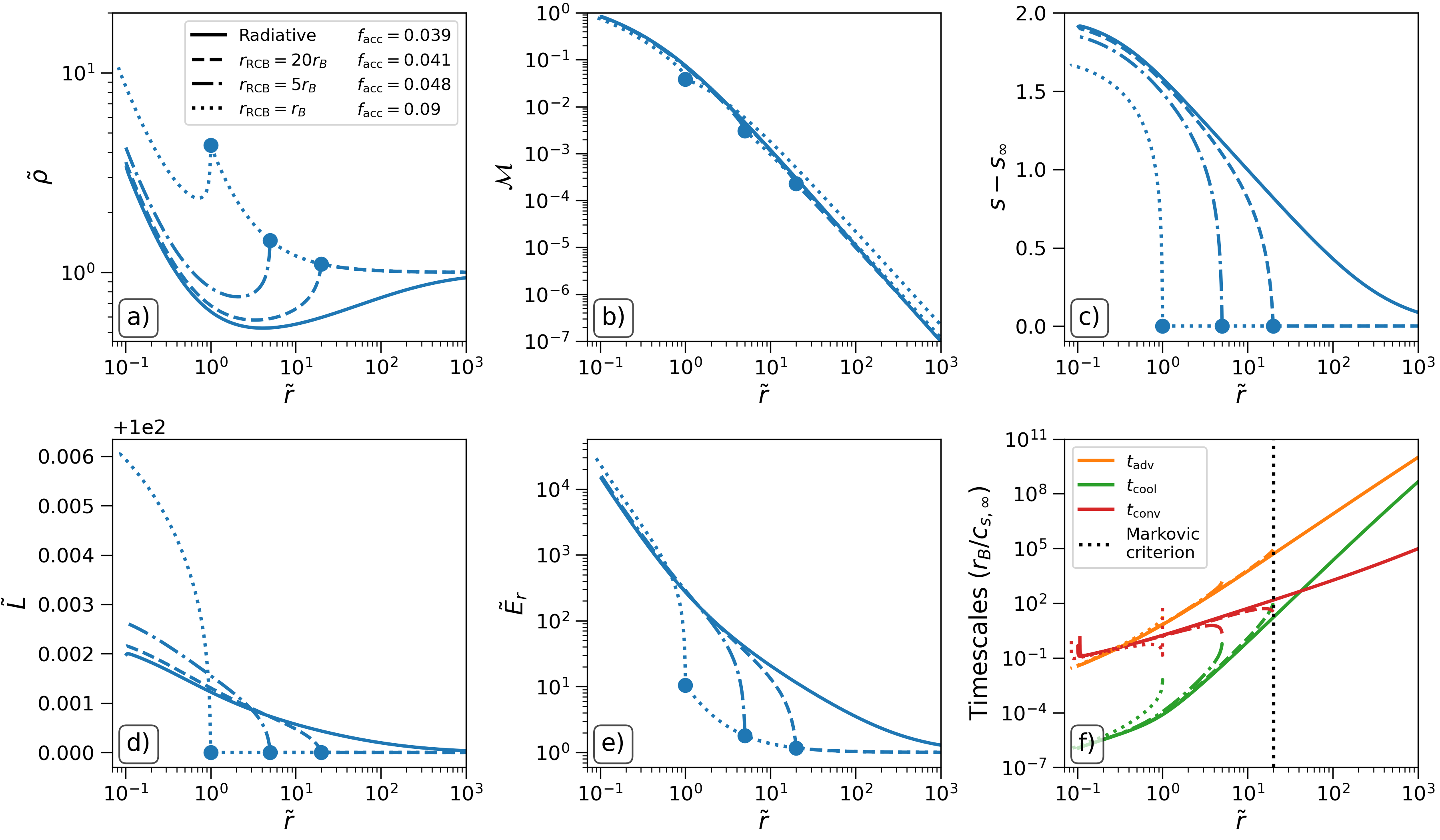}
    \caption{Comparison of a fully radiative model (solid) versus radiative-convective models with $r_{\rm RCB}=20r_B$ (dashed) and $r_{\rm RCB}=5r_B$ (dot-dash) and $r_{\rm RCB}=r_B$ (dotted). All models set $\beta=10^{-3}$, $\tau_B=1$, $\tilde{L}_\infty=100$. Panels (a-e) display profiles of the dependent variables as in Figures \ref{fig:thin} \& \ref{fig:thick}. Markers are placed at the location of the radiative-convective boundary. Panel (f): the dimensionless timescales associated with each model. The vertical line shows the RCB $(r_{\rm RCB}=20 r_B)$ expected from equation (\ref{eq:markcrit}). This is only a factor of $\approx 3$ interior to the boundary estimated by the more approximate condition $t_{\rm conv}< t_{\rm adv},\, t_{\rm cool}$.}
    \label{fig:rcb}
\end{figure*}

In Figure \ref{fig:rcb}, we apply this radiative-convective  procedure to a model with $\beta=10^{-3}$, $\tau_B=1$, and $\tilde{L}_\infty=100$. The fully radiative model has   suppressed accretion with $f_{\rm acc}=0.039$ and the Markovi\'{c} criterion predicts an RCB at $r_{\rm RCB}=20r_B$. We test several values of $r_{\rm RCB}$ including $20r_B$ and lower values, to study the effect of RCB location. Compared to the fully radiative model, we see that the radiative-convective models exhibit cusps at the  transition from adiabatic to radiative flow. These discontinuous derivatives in the flow variable would be smoothed in a more complete convective model. 

In this model, including outer convective zones increases the accretion rate, i.e.\ reduces the effect of radiative suppression.  However this effect is small when the convective zone is outside $r_B$.  Specifically $f_{\rm acc}$ increases by $\sim 5\%$ and $\sim25\%$ for $r_{\rm RCB} = 20 r_B$ and $5r_B$, respectively.  When $r_{\rm RCB} = r_B$, the effect is more significant, as $f_{\rm acc}$ more than doubles (but suppression is still significant with $f_{\rm acc} < 0.1$), and the higher density and lower entropy at the sonic point is evident in  Figure \ref{fig:rcb}.

Since this simplified model overestimates the effects of convection, we draw the following conclusions.  When the RCB in radiative models is outside $r_B$, the effect of including convection ranges from negligible to an order unity correction.  When this RCB is inside $r_B$, a more detailed convective model is needed to more accurately model the radiative suppression of accretion rates.  To consider when these corrections might be significant, the results of Figure \ref{fig:rcb} and the parameter values of equations (\ref{eq:taubeta_vals}) show that these convective corrections are unlikely to be significant for most planet formation applications.  The reasoning is that $\beta$ values should generally be small, and where they are larger (the outer disk), $\tau_B$ values are lower.  Figure \ref{fig:rcb} shows that these parameter values give non-existent or distant outer convective zones in accretion flows.

\subsection{Grey Opacity}
The grey opacity adopted here potentially suffers when modeling optically thin transfer. Photons released from a photosphere at $\tilde{r}\approx \tau_B^2$ have photon energy $h\nu\sim k_B T_{\rm phot}$. As they free-stream, the radiation energy density/temperature drops via geometrical dilution, but photon energies remain fixed. Grey emission/absorption $\sim\kappa (a_rT^4-E_r)$ is degenerate in that a dilute field from a hot source looks identical to a stronger field from a cool one.

To break this degeneracy, it is appropriate to split the emission/absorption opacities $\sim \kappa(T_{\rm phot})E_r - \kappa(T)a_rT^4$. The energy density $E_r$ retains its geometric dilution, but absorption operates at the appropriate $T_{\rm phot}$ energy/frequency scale. In this case, radiative equilibrium is given by $a_rT^4=\xi E_r$ with $\xi\equiv\kappa(T_{\rm phot})/\kappa(T)$ and the gas superheats by $\xi^{1/4}$ relative to the radiation temperature $(E_r/a_r)^{1/4}$. Taking $\kappa\sim T^n$, then $\xi=(T_{\rm phot}/T)^n$ and $a_rT^{4+n}=T_{\rm phot}^nE_r$. With $E_r\sim r^{-2}$, the temperature profile changes slope to $T\sim r^{-2/(4+n)}$, shallower than the necessary $r^{-1}$ for constant luminosity so long as $n>-2$. Repeating the analysis of \S \ref{sec:thin}, we have a dimensionless temperature law of $\tilde{T}^{4+n}= \tilde{T}_{\rm phot}^n\tilde{E}_r= \tilde{T}_{\rm phot}^n \tilde{L}_\infty \tilde{r}^{-2}$. The relevant scale is defined by $GM/r_L=\mathcal{R}T_L$ so $\tilde{r}_L\simeq \tilde{T}_L^{-1}$. Evaluating the temperature law at $\tilde{r}_L$, gives $\tilde{T}_L$
\begin{equation}
\begin{split}
 \tilde{T}_L\simeq\left(\tilde{T}_{\rm phot}^{n}\tilde{L}_\infty\right)^{1/(2+n)}.
\end{split}
\end{equation}
Applying pressure equilibrium $\tilde{\rho}_L\simeq 1/\tilde{T}_L$ one has $\dot{M}/\dot{M}_{\rm ad}\simeq \tilde{\rho}_L\tilde{c}_L\tilde{r}_L^2\simeq T_L^{-5/2}$ or 
\begin{equation}
    f_{\rm acc}\simeq (\tilde{T}_{\rm phot}^n \tilde{L}_\infty)^{-5/(4+2n)}
\end{equation}
We see that setting $n=1$, recovers the standard superheated temperature profile $\tilde{T}\sim \tilde{r}^{-2/5}$ \citep{ChiangGoldreich1997} and gives scalings $\tilde{T}_L\sim (\tilde{T}_{\rm phot}\tilde{L}_\infty)^{1/3}$, $f_{\rm acc}\sim (\tilde{T}_{\rm phot}\tilde{L}_\infty)^{-5/6}$. For $n>0$ in general, one finds that the slope of this luminosity suppression which was $-5/4$ is flattened to $-5/(4+2n)$.

However, the net effect is only to suppress accretion relative to grey opacities, so that our grey models are an upper limit on the accretion rate. To see this, note that the characteristic grey temperature scale of \S\ref{sec:thin} was $\tilde{T}_{L,\rm grey}\simeq \tilde{L}_\infty^{1/2}$ so here, $\tilde{T}_L \simeq \tilde{T}_{\rm phot}^{n/(2+n)}\tilde{T}_L^{2/(2+n)}$ is a weighted geometric mean. This implies $\tilde{T}_{L,\rm grey}<\tilde{T}_{L} < \tilde{T}_{\rm phot}$ so $f_{\rm acc} \simeq T_L^{-5/2}< \tilde{T}_{L,\rm grey}^{-5/2}\simeq f_{\rm acc, grey}$ and accretion is only further reduced. Of course, all of this requires $\kappa(T_{\rm phot}) > \kappa(T)$, but this is guaranteed for any dust opacity \citep{Malygin+2014, Birnstiel+2018}.

\section{Conclusion}\label{sec:end}
Having investigated the problem of steady-state, spherically-symmetric radiative accretion for gas-pressure dominant regimes, there emerges rich physics not seen in other previously studied regimes. The main results of this work are summarized as:
\begin{enumerate}
    \item Radiative feedback can suppress accretion by heating the accreting matter and changing the location of the sonic point. The modification is most extreme for large optical depths $\tau_B$, large luminosities $\tilde{L}_\infty$, and/or long cooling times $\beta$, and can be orders-of-magnitude in effect (\S \ref{sec:num}).
    \item In the limit of high luminosity, radiative suppression scales inversely with luminosity $f_{\rm acc}\sim \tilde{L}_\infty^{-5/4}$, independent of opacity \& cooling time (\S \ref{sec:thin}).
    \item  In the limit of high optical depth, radiative suppression scales as $f_{\rm acc}\sim \tau_B^{-10/11}\beta^{-5/11}$, independent of luminosity. However, these solutions with radially varying luminosity do not admit self-consistent freefall solutions in the interior making them unlikely to apply to astrophysical systems. 
    \item In the intermediary limit of both high optical depth and high luminosity, radiative suppression scales jointly as $f_{\rm acc}\sim(\tilde{L}_\infty\tau_B)^{-5/8}$ (\S \ref{sec:thick}), independent of cooling time (\S \ref{sec:thick}).
    \item Radiative suppression is coincident with adverse entropy gradients which would be Schwarzschild unstable. In practice however, rapid cooling can make convection inefficient and accretion rates more radiative (\S \ref{sec:convect}).
\end{enumerate}

While this work has been constructed to be mostly agnostic about the underlying type of astrophysical environment, follow-up work will be focused on dedicated astrophysical applications. Perhaps the most relevant application for these models, as we have alluded to, is the relatively cold $\sim 10^2$ K regime of protoplanetary environments. In the context of planet formation, specifically core accretion, for planets more massive than $\sim 20M_\oplus$, self-gravity drives a hydrodynamic phase of so-called runaway growth. As a result, population synthesis models \citep{Emsenhuber+2021} sometimes adopt an adiabatic Bondi rate to parameterize this accretion process. However, the intrinsic luminosity of the hot young planet or the accretion luminosity can be substantial and models derived here which include some radiative feedback are preferable. Taking these 1D hydrodynamic accretion rates with radiative feedback, and extending them to realistic non-constant opacities for population synthesis calculations is the focus of our follow-up Paper II. This first subsequent paper looks in detail at the planet-formation context and answers some of the uncertainties raised regarding convection and accretor luminosity by running direct numerical simulations. In that case, the radiative accretion rates presented here appear to be more-or-less correct. 

Protoplanetary environments may not be the only applicable parameter space for these radiative feedback models. In AGN disks, the thermodynamics are such that radiation pressure tends to be dominant over the thermal pressure, but only moderately \citep{SirkoGoodman2003}. Given the range of possible parameters and the large physical extent of the disk (from Schwarzschild radius to parsec), there are potential conditions or locations where the gas may become thermally dominated and in fact, the fiducial models of \citep{Thompson+2005} show some evidence of this. Under the right conditions, for accretion onto stars or black holes in these disks, these models could be of some interest.

Wind accretion in stellar binaries is another potential application of these models. Current 1D models, tend to adopt an adiabatic Bondi-Hoyle like accretion rate when modeling the evolution of a star accreting from a companion wind. In this context, the luminosity would not be powered by accretion but rather by the intrinsic luminosity of the accreting star. Thus given optical and thermal properties of the donor star's wind, a revised accretion rate including the feedback of the luminous accretor may be calculated via the framework here. For parameters which place the flow in a thermal-pressure dominated regime, existing models for highly luminous accretors could be over-estimating the prescribed accretion rate by not including the radiative feedback effects found here.

For star formation applications, the accretion suppression mechanism is possible relevance. For molecular cloud conditions, we see that generally the optical depth is low $\tau_B \ll 1$
\begin{equation}
\begin{split}
    \tau_B\approx 0.02\left(\frac{\rho_\infty}{10^{-19}\text{ g/cm}^3}\right)\left(\frac{\kappa}{1\text{ cm}^2/{\rm g}}\right)\\
    \times\left(\frac{M}{1 M_\odot}\right)\left(\frac{T_\infty}{10 \text{ K}}\right)^{-1}
\end{split}
\end{equation}
In this case $\tilde{L}_\infty < 1$ gives isothermal accretion while $\tilde{L}_\infty > 1$ gives optically thin suppression $f_{\rm acc}\sim \tilde{L}_\infty^{-5/4}$. The luminosity at which systems fall into the thin suppression regime is: 
\begin{equation}
    L_\infty \gtrsim L_B = 100 L_\odot\left(\frac{M}{1M_\odot}\right)^2\left(\frac{T_\infty}{10\text{ K}}\right)^2
\end{equation} 
requiring a fairly bright source. In terms of a mass accretion rate $L\approx GM\dot{M}/r$, this is
\begin{equation}
    \dot{M}\gtrsim 3\times 10^{-6} M_\odot /{}\rm yr
\left(\frac{M}{1M_\odot}\right)\left(\frac{T_\infty}{10 \text{ K}}\right)^2\left(\frac{r_{\rm src}}{1 R_\odot}\right)
\end{equation}
In protostellar applications \citep{Stahler+1980}, there is additional substructure to the flow, like dust sublimation regions in the interior. We would not expect these to change the results so much since these features occur well interior to the sonic point. The exception is if the temperature profile becomes steeper than $1/r$ in which case our constant luminosity approximation could fail. The greater barrier is in the inclusion of self-gravity, which is not treated here.

Ultimately we recommend follow-up work similar to our Paper II be undertaken for these other potential applications of interest. For the present moment however, we simply provide the radiative accretion rates (through a downloadable table or the analytic formula of Appendix \ref{app:formula}) such that they may be tested against existing Bondi accretion parameterizations which use only an adiabatic rate.

\begin{acknowledgments}
We thank Zhaohuan Zhu for useful discussions.  This work is supported by the National Aeronautics and Space Administration under Agreement No. 80NSSC21K0593 for the program
“Alien Earths.” This work has also been supported by National Aeronautics and Space Administration under Agreement No. 80NSSC24K0163. This material is based upon High Performance Computing (HPC) resources supported by the University of Arizona TRIF, UITS, and Research, Innovation, and Impact (RII) and maintained by the UArizona Research Technologies department. Resources supporting this work were also provided by the NASA High-End Computing (HEC) Program through the NASA Advanced Supercomputing (NAS) Division at Ames Research Center.
\end{acknowledgments}

\section*{Data Availability Statement}
Tabulated data for $f_{\rm acc}$ and other tools are available at https://github.com/apbailey/radiative-bondi-products. Additional data underlying this article will be shared on reasonable request to the corresponding author.

\begin{contribution}
AB led the production of this work -- development of numerical solutions, analysis of the models, and writing of the manuscript. AY motivated this work with initial development of a set of radiating Bondi equations, contributed to the writing, and constructed many of the analytic arguments for Section \ref{sec:anal}. KK provided substantial guidance and feedback throughout the entire process.
\end{contribution}

%

\software{This work made extensive use of the SciPy \citep{2020SciPy}, Matplotlib \citep{Matplotlib}, and NumPy \citep{Numpy} packages.}

\bibliography{sample7}{}
\bibliographystyle{aasjournalv7}


\appendix

\section{Integration Procedure at Large Radius}\label{app:inteq}
As mentioned in Section \ref{sec:steady}, directly integrating the system of equations (\ref{eq:dimless}) poses numerical challenges at large radius.  The validity of the radiative equilibrium approximation at large radii (see \S\ref{sec:radeq}) means that small numerical errors in $T^4-E_r$ drives errors in the luminosity, via equation (\ref{eq:dim-L}).  This problem is caused by the need to use a large radius outer boundary for large luminosities, optical depths and/or cooling times (\S\ref{sec:bc}).  The problem is not solved by using integrators designed for stiff ODEs.

We address this problem by solving a separate radiative equilibrium system of equations at large radius and then switching to the full radiative disequilibrium system at smaller radii.  As described in \S\ref{sec:radeq}, radiative equilibrium sets $\tilde{E}_r=\tilde{T}^4$ and replaces equation (\ref{eq:dim-er}) with equation (\ref{eq:reqT}). 

For further numerical simplification, we use the constancy of $f_{\rm acc}$ in equation (\ref{eq:facc}) to eliminate $d \ln\mathcal{M}/dr + (1/2) ds/dr$ between equations (\ref{eq:dim-rho}) \& (\ref{eq:dim-M}) to give
\begin{equation}\label{eq:app3}
    \frac{d\tilde{\rho}}{d\ln \tilde{r}} = \frac{\tilde{\rho}}{\gamma \mathcal{M}^2 - 1}\left(-2\gamma\mathcal{M}^2 + \frac{2\gamma}{\tilde{T}\tilde{r}}+\frac{1}{\tilde{T}}\frac{d\tilde{T}}{d\ln \tilde{r}}\right).
\end{equation}

For the energy/luminosity equation we use (\ref{eq:dim-s}), swapping entropy for $\tilde{\rho}$, $\tilde{T}$:
\begin{equation}\label{eq:app1}
    \partial_{\tilde{r}} \tilde{L} = 4q_\gamma f_{\rm acc} \beta\tau_B \tilde{T}\left(\partial_{\tilde{r}}\ln \tilde{T} - (\gamma-1)\partial_{\tilde{r}}\ln\tilde{\rho}\right).
\end{equation}
However, since equations (\ref{eq:reqT}), (\ref{eq:app1}), (\ref{eq:app3}) couple $dT/dr$, $dL/dr$, and $d\rho/dr$ implicitly, we substitute equations (\ref{eq:app3}) \& (\ref{eq:reqT}) into equation (\ref{eq:app1}) to obtain an explicit luminosity equation
\begin{equation}\label{eq:app4}
    \frac{d\tilde{L}}{d\ln \tilde{r}}= \frac{4q_\gamma f_{\rm acc}\beta\tau_B}{1 + C_3}\left[ \frac{\gamma-1}{\gamma \mathcal{M}^2 - 1}\left(2\gamma\mathcal{M}^2\tilde{T} - \frac{2\gamma}{\tilde{r}}\right)-\left(\frac{\mathcal{M}^2 - 1}{\gamma \mathcal{M}^2 - 1} \right)\frac{{\gamma \tilde{L}}}{4\tilde{T}^3\tilde{r}^2}\left(2  + 3\tilde{\rho} \tau_B \tilde{r} \right)\right],
\end{equation}
\begin{equation}
    C_3 \equiv \frac{{2\gamma q_\gamma f_{\rm acc}\beta\tau_B}}{\tilde{T}^3\tilde{r}^2}\left(\frac{\mathcal{M}^2 - 1}{\gamma \mathcal{M}^2 - 1} \right),
\end{equation}
suitable for integration by general-purpose initial-value problem solvers. The set of equations (\ref{eq:facc}), (\ref{eq:reqT}), (\ref{eq:app3}), (\ref{eq:app4}), forms a closed system of equilibrium equations for the variables $(\mathcal{M},  \tilde{T}, \tilde{\rho}, \tilde{L})$ that we solve numerically. To determine when to switch to the disequilibrium system, at each step in the integration we assess the ``poorness" of our radiative equilibrium assumption from the luminosity gradient $|\delta E/E|\equiv|\partial_{\tilde{r}}\tilde{L}|/(\tilde{r}^2\tilde{\rho}\tilde{T}^4\tau_B)\sim|\tilde{T}^4-\tilde{E}_r|/\tilde{T}^4$. In the numerical solutions of Section \ref{sec:models}, we switch to the disequilibrium system when $|\delta E/E| > 10^{-3}$ or $r<10^2r_B$, whichever happens first. The initial conditions on the disequilibrium system are then set from the equilibrium solutions at the switching point -- the only potentially ambiguous one is $\tilde{E}_r$ which is simply set to $\tilde{T}^4$ and then quickly evolves to the appropriate level of disequilibrium.

\section{Non-Singular Expressions for Integration Near the Sonic Point}\label{app:quadratic}
Here we derive an alternate, non-singular expression for $\partial_{\tilde{r}}\ln\mathcal{M}$ suitable for explicit integration through the sonic point. Noting that Eq.\ (\ref{eq:dim-M}) is of the form $\partial_{\tilde r}\ln\mathcal{M} = \mathcal{N}/\mathcal{D} + \mathcal{R}$
\begin{equation}\label{eq:ndr}
  \mathcal{N} = \frac{1}{\tilde{r}} - \frac{1}{\tilde{T}\tilde{r}^2} - \frac{ \partial_{\tilde{r}}s}{2\gamma},\quad
  \mathcal{D} = \frac{\mathcal{M}^2 - 1}{\gamma+1},\quad
  \mathcal{R} = \frac{\gamma -1}{\tilde{r}} - \frac{\partial_{\tilde{r}}s}{2},
\end{equation}
we take the limit $r\rightarrow r_\mathcal{M}$ $(\mathcal{M}\rightarrow 1)$ and leverage regularity at the sonic point ($\mathcal{N}/\mathcal{D}\rightarrow 0/0$) to apply L'H\^{o}pital's rule to $\mathcal{N}/\mathcal{D}$ giving
\begin{equation}\label{eq:quad}
    \left.\partial_{\tilde{r}}\mathcal{N}\right|_{\mathcal{M}\rightarrow1}=\left[\partial_{\tilde{r}}\mathcal{D}(\partial_{\tilde{r}}\ln\mathcal{M} - \mathcal{R})\right]_{\mathcal{M}\rightarrow1}.
\end{equation}
Because the system (\ref{eq:dim-rho})-(\ref{eq:dim-er}) is linear in the derivatives $\partial_{\tilde{r}}$, then $\mathcal{R}$ is at most also linear in $\partial_{\tilde{r}} \ln\mathcal{M}$. Similarly, the chain rule applied to $\mathcal{N}$ gives a $\partial_{\tilde{r}}\mathcal{N}$ which is at most linear in $\partial_{\tilde{r}}\ln\mathcal{M}$. Combined with the linearity of $\partial_{\tilde{r}}\mathcal{D} = 2\mathcal{M}^2\partial_{\tilde{r}}\ln\mathcal{M}/(\gamma+1)$, Eq.\ (\ref{eq:quad}) is a quadratic $q_2(\partial_{\tilde{r}}\ln\mathcal{M})^2 + q_1 (\partial_{\tilde{r}}\ln\mathcal{M}) + q_0 = 0$. The goal here is algebraic expansion of the terms in Eq.\ (\ref{eq:quad}) to write the quadratic coefficients $q_i$ in terms of state variables $(\tilde{r}, \tilde{\rho}, \mathcal{M}, s, \tilde{L}, \tilde{E}_r)$ -- the only non-trivial one being $\partial_{\tilde{r}}\mathcal{N}$. Differentiating $\mathcal{N}$ gives
\begin{equation}\label{eq:dN}
    \partial_{\tilde{r}}\mathcal{N} = -\frac{1}{\tilde{r}^2} + \frac{2}{\tilde{T} \tilde{r}^3} +\frac{\partial_{\tilde{r}}\ln \tilde{T}}{\tilde{T} \tilde{r}^2} - \frac{\partial^2_{\tilde{r}}s}{2\gamma}.
\end{equation}
For the second-order entropy derivative in Eq.\ (\ref{eq:dN}) we use Eq. (\ref{eq:dim-s}) to write:
\begin{equation}\label{eq:d2s}
    \partial_{\tilde{r}}^2 s  = \partial_{\tilde{r}}s
\left(\frac{\partial_{\tilde{r}}^2 \tilde{L}}{\partial_{\tilde{r}}\tilde{L}} -\partial_{\tilde{r}}\ln\tilde{T}\right)
\end{equation}
and further reduce it to first-order derivatives via differentiation of Eq.\ (\ref{eq:dim-L})
\begin{equation}\label{eq:d2L}
      \frac{\partial_{\tilde r}^2\tilde L}{\partial_{\tilde r}\tilde L} = \frac{2}{\tilde r}
  + \partial_{\tilde r}\ln\tilde\rho
  + \frac{4\tilde T^4\partial_{\tilde r}\ln\tilde T
  - \partial_{\tilde r}\tilde E_r}{\tilde{T}^4 - \tilde{E}_r}.
\end{equation}
All that remains is substitution of Eqs.\ (\ref{eq:d2L})$\rightarrow$(\ref{eq:d2s})$\rightarrow$(\ref{eq:dN})$\rightarrow$(\ref{eq:quad}) and rewriting of $\partial_{\tilde{r}}\ln \tilde{T}$, $\partial_{\tilde{r}}\ln\tilde{\rho}$ in terms of state variables. We use the following expressions derived via the equation of state and mass continuity (\ref{eq:dim-rho}) to do so:
\begin{equation}
    \partial_{\tilde{r}}\ln\tilde{T}  = -\frac{4\mathcal{R}}{\gamma+1} - \frac{2(\gamma-1)}{\gamma+1}\partial_{\tilde{r}} \ln \mathcal{M} 
\end{equation}
\begin{equation}    
   \partial_{\tilde r}\ln\tilde\rho=
    \frac{2}{\gamma+1}\left(\mathcal{R} - \partial_{\tilde{r}} \ln \mathcal{M} \right) - \frac{2}{\tilde{r}}
\end{equation}
After substitution and collecting powers of $\partial_{\tilde{r}}\ln\mathcal{M}$ in Eq.\ (\ref{eq:quad}), some additional algebra yields the quadratic coefficients:
\begin{align}
q_2 &= \mathcal M^2\\
q_1 &= -\mathcal M^2\mathcal R + \frac{\gamma-1}{\tilde T\tilde r^2} - \frac{(2-\gamma)\partial_{\tilde r}s}{2\gamma} - \frac{(\gamma-1)\tilde T^{5/2}}{2\gamma\beta\mathcal M}\\
q_0 &= (\gamma+1)\left[\frac{1}{2\tilde r^2}-\frac{1}{\tilde T\tilde r^3}-\frac{\partial_{\tilde r}\tilde E_r}{16\gamma\beta\tilde T^{3/2}\mathcal M}\right] + \mathcal R\left[\frac{2}{\tilde T\tilde r^2}+\frac{3\partial_{\tilde r}s}{2\gamma}-\frac{\tilde T^{5/2}}{\gamma\beta\mathcal M}\right]
\end{align}
with any remaining $\partial_{\tilde{r}}s$, $\partial_{\tilde{r}}\tilde{E}_r$, $\mathcal{R}$ trivially converted to state variables via Eqs.\ (\ref{eq:dim-s})-(\ref{eq:dim-er}). Near the sonic point, we simply replace Eq.\ (\ref{eq:dim-M}) with 
\begin{equation}
\partial_{\tilde r}\ln\mathcal{M} = \frac{-q_1 - \sqrt{q_1^2 - 4q_0q_2}   }{2q_0}
\end{equation}
which holds exactly at $r_{\mathcal{M}}$ and approximately in its vicinity. In this case, the negative root is selected, corresponding to an accretion solution.

\section{Analytic Formula for Radiative Bondi Accretion Rate}\label{app:formula}
Here we use the numerical solutions to refine normalizations for the derived scalings compiled in Section \ref{sec:unify}. This allows us to construct a well-fitting analytic formula for the mass accretion rate $f_{\rm acc}(\tau_B,\tilde{L}_\infty, \beta)$ when $\beta\lesssim 1$. 

We begin by considering the nearly isothermal solutions. These solutions have $f_{\rm acc}\approx 1$ and are expected occur when $\tau_B\beta\lesssim\tilde{L}_\infty\lesssim \min(1,\tau_B^{-1})$ -- that is, defined by the boundaries $\tilde{L}_\infty/(\tau_B\beta)=1$, $\tilde{L}_\infty\tau_B=1$, and $\tilde{L}_\infty=1$. The first boundary separates isothermal solutions from ``no-solution" regions, the second from optically thick $(\tilde{L}_\infty \tau_B) \gg1$ regions, and the third from optically thin regions. However, as stated, these boundaries are only estimates and can be seen to differ from the numerical solution by a factor $\approx 10$. For a more accurate prescription we prefer numerically estimated values for these factors. This prompts us to instead say the isothermal region $f_{\rm acc}\approx 1$ is defined by $\tilde{L}_\infty/(\tau_B\beta)=a_1$, $\tilde{L}_\infty\tau_B=a_2$, and $\tilde{L}_\infty=a_3$, with Figure \ref{fig:facc} suggesting better normalizations $a_1\approx a_3\approx 10$ and $a_2\approx 1$. 

We now consider optically thin solutions $f_{\rm acc}\sim \tilde{L}_\infty^{-5/4}$ occurring for $\tau_B\ll 1$. $\tilde{L}_\infty=a_3$ now defines their boundary from isothermal solutions and thus to enforce continuity of $f_{\rm acc}$ into the isothermal regime, we require that $f_{\rm acc}=(\tilde{L}_\infty/a_3)^{-5/4}$ in optically thin regions. Of course, this solution may not continue to arbitrarily high $\tau_B$, and at $(\tilde{L}_\infty, \tau_B)=(a_3, a_2/a_3)$, the isothermal boundary deviates from being $\tau_B$-independent. This gives a point from which to construct an upper $\tau_B$-boundary on our optically thin solutions. The remainder of this boundary comes from requiring $f_{\rm acc}$ be continuous in the transition from optically thin to optically thick $f_{\rm acc}\sim (\tilde{L}_\infty\tau_B)^{-5/8}\sim\tilde{L}_\infty^{-5/4}$ or $\tau_B\sim \tilde{L}_\infty$. Thus the $\tau_B$-boundary on optically thin solutions is given by $\tau_B<a_2\tilde{L}_\infty/a^2_3$. We combine this with the isothermal boundary to give single condition for optically thin regimes $\tilde{L}_\infty >\max(a_3, a_3^2\tau_B/a_2)$.

Optically thick solutions lie between no-solution and optically thin regimes, giving the joint constraint $a_2/\tau_B<\tilde{L}_\infty<a_2\tau_B/a_3^2$. We subdivide these into the high luminosity ``Case 1" and low luminosity ``Case 2" regimes of \S\ref{sec:thick}.  Similar to Section \ref{sec:unify}, where a boundary was defined with fixed point $(\tilde{L}_\infty, \tau_B) = (\beta^{1/2}, \beta^{-1/2})$ (i.e.\ the intersection of $\tilde{L}_\infty\tau_B=1$ and $\tilde{L}_\infty/(\tau_B\beta)=1$) and requiring continuity of $f_{\rm acc}$, here we use fixed point $(\tilde{L}_\infty, \tau_B) = (\sqrt{a_1a_2\beta}, \sqrt{a_2/(a_1\beta)})$. This gives a boundary $\tilde{L}_\infty=(\tau_B/a_2)^{5/11}(a_1a_2\beta)^{8/11}$. At high luminosity solutions must match onto isothermal solutions along the boundary $\tilde{L}_\infty=a_2/\tau_B$ giving $f_{\rm acc}=(\tilde{L}_\infty\tau_B/a_2)^{-5/8}$. At low luminosity the accretion rate is $f_{\rm acc} = (a_1\beta\tau_B^2/a_2)^{-5/11}$ from requiring $f_{\rm acc}= 1$ at the fixed point. 

\begin{figure*}
    \centering
    \includegraphics[width=\linewidth]{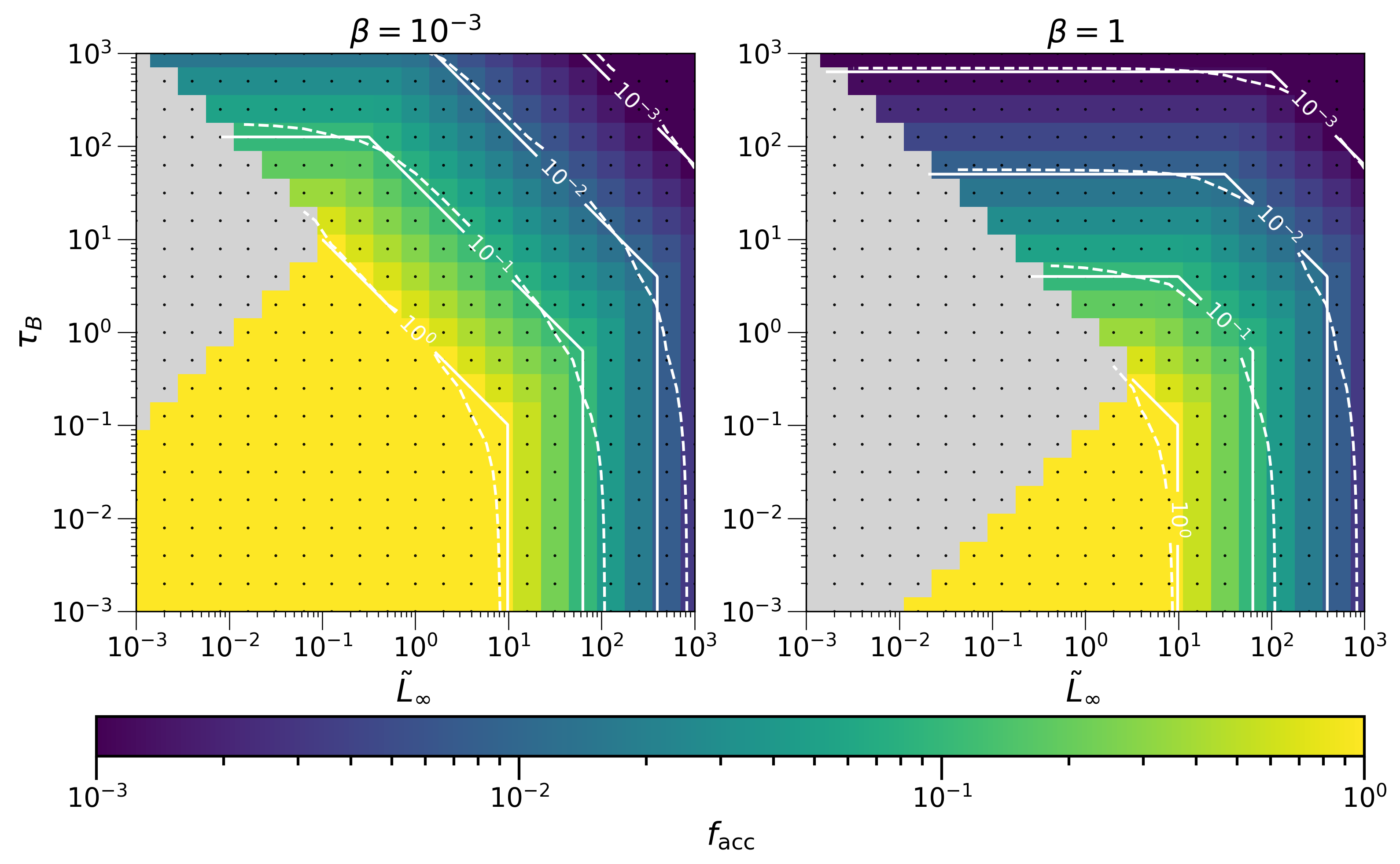}
    \caption{Maps of $f_{\rm acc}$ constructed in the style of Figure \ref{fig:facc}, but instead of using the true numerical solutions, $f_{\rm acc}$ for each point is colored according to our analytic expression (\ref{eq:analytic-facc}) (with $a_1=a_3=10$, $a_2=1$). Solid contours are placed at the $f_{\rm acc}=[10^{-4}, 10^{-3}, 10^{-2}, 10^{-1}, 10^{0}]$ levels with the dashed lines coming from the numerical solutions of Figure \ref{fig:facc} and solid lines coming from our analytic formula. For $\beta=10^3$, we plot only the optically thick regime of solutions because our unified equation (\ref{eq:analytic-facc}) is only strictly applicable for $\beta\lesssim1$.}
    \label{fig:analytic-facc}
\end{figure*}

In total, we combine the regimes into a piecewise formula for $f_{\rm acc}$ which is both continuous and agrees well with the numerical solutions:
\begin{equation}\label{eq:analytic-facc}
\begin{split}
    f_{\rm acc} = 
    \begin{cases}
    \text{No solution} &  \tilde{L}_\infty/(a_1\beta)<\tau_B < a_2\tilde{L}_\infty^{-1}\\
    1 & a_1\tau_B\beta <\tilde{L}_\infty < \min(a_3, a_2\tau_B^{-1})\\
    (\tilde{L}_\infty/a_3)^{-5/4} & \tilde{L}_\infty >\max(a_3, a_3^2\tau_B/a_2)\\
    f_{\rm thick} & a_2/\tau_B<\tilde{L}_\infty<a_3^2\tau_B/a_2
    \end{cases}\\
    f_{\rm thick} = 
    \begin{cases}
    (\tilde{L}_\infty\tau_B/a_2)^{-5/8} & \tilde{L}_\infty > (\tau_B/a_2)^{5/11}(a_1a_2\beta)^{8/11}\\
    (a_1\beta\tau_B^2/a_2)^{-5/11}& \tilde{L}_\infty <(\tau_B/a_2)^{5/11}(a_1a_2\beta)^{8/11}
    \end{cases}
\end{split}
\end{equation}
where our preferred values are $a_1\approx a_3\approx 10$ and $a_2\approx 1$. It is seen that by setting $a_1=a_2=a_3=1$ this formula reduces to the appropriate scalings and regimes given in Section \ref{sec:unify}. As a check on the accuracy of this formula, we compute $f_{\rm acc}$ according to equation (\ref{eq:analytic-facc}) across the parameter space of our numerical models. In Figure \ref{fig:analytic-facc}, we plot the analytic $f_{\rm acc}$ in the style of Figure \ref{fig:facc}. We include contours of $f_{\rm acc}$ for both the analytic (dashed) and numerical solutions (solid) and find very good agreement in the $\beta\leq1$ regimes where radiative equilibrium is a good approximation. For $\beta>1$, disequilibrium effects creep into the parameter space and new regimes which we have not sufficiently described analytically are introduced. Since equation (\ref{eq:analytic-facc}) was designed with radiative equilibrium and $\beta\lesssim1$ in mind, applying equation (\ref{eq:analytic-facc}) to $\beta>1$ results in overlapping boundaries between different regimes and overall inconsistency so $\beta=10^3$ is not plotted in Figure \ref{fig:analytic-facc}. 

\end{document}